\documentclass[usenatbib]{mn2e}
\usepackage{natbib}
\usepackage{amssymb}
\usepackage{epsfig}
\usepackage{rotating}
\usepackage{aas_macros}
\usepackage{url}
\usepackage{amsmath}
\usepackage{multirow}
\usepackage{fixltx2e}
\usepackage{color}

\newcommand{\Msol}{\ensuremath{\mathrm{M_{\odot}}}}

\newcommand{\rf}{\ensuremath{R_{\mathrm{500}}}}

\newcommand{\fgas}{\ensuremath{f_{\mathrm{gas}}}}
\newcommand{\mgas}{\ensuremath{M_{\mathrm{gas}}}}

\newcommand{\Mgas}{\ensuremath{M_{\mathrm{gas}}}}
\newcommand{\rhog}{\ensuremath{\rho_{\mathrm{gas}}}}

\newcommand{\lik}{\ensuremath{{\cal L}}}

\newcommand{\Chandra}{\emph{Chandra}}

\newcommand{\XMM}{\emph{XMM-Newton}}

\newcommand{\chisq}{\ensuremath{\chi^2}}

\newcommand{\gta}{\,\rlap{\raise 0.4ex\hbox{$>$}}{\lower 0.6ex\hbox{$\sim$}}\,}
\newcommand{\lta}{\,\rlap{\raise 0.4ex\hbox{$<$}}{\lower 0.6ex\hbox{$\sim$}}\,}

\newcommand{\km}{\mbox{\ensuremath{\mathrm{~km}}}}

\newcommand{\kpc}{\mbox{\ensuremath{\mathrm{~kpc}}}}
\newcommand{\Mpc}{\mbox{\ensuremath{\mathrm{~Mpc}}}}
\newcommand{\s}{\mbox{\ensuremath{\mathrm{~s}}}}

\newcommand{\keV}{\mbox{\ensuremath{\mathrm{~keV}}}}
\newcommand{\erg}{\mbox{\ensuremath{\mathrm{~erg}}}}

\newcommand{\muT}{\mbox{\ensuremath{~\mathrm{\muT}}}}

\newcommand{\pMpc}{\ensuremath{\mathrm{\Mpc^{-1}}}}
\newcommand{\ps}{\ensuremath{\mathrm{\s^{-1}}}}

\newcommand{\ergps}{\ensuremath{\mathrm{\erg \ps}}}

\newcommand{\kmpspMpc}{\ensuremath{\mathrm{\km \ps \pMpc\,}}}

\newcommand{\YM}{\mbox{\ensuremath{\mathrm{Y_{X}-M_{500}}}}}

\newcommand{\LCDM}{$\Lambda$CDM~}

\renewcommand{\mgas}{\ensuremath{\mu}}
\newcommand{\mobs}{\ensuremath{M_{\mathrm{obs}}}}
\newcommand{\mfit}{\ensuremath{M_{\mathrm{fit}}}}
\newcommand{\sigm}{\ensuremath{\sigma_M}}
\newcommand{\tobs}{\ensuremath{T_{\mathrm{obs}}}}
\newcommand{\sigt}{\ensuremath{\sigma_T}}
\newcommand{\tmod}{\ensuremath{T_{\mathrm{mod}}}}
\newcommand{\tint}{\ensuremath{T_{\mathrm{int}}}}

\newcommand{\mgobs}{\ensuremath{\mu_{\mathrm{obs}}}}
\newcommand{\sigmg}{\ensuremath{\sigma_\mu}}
\newcommand{\mgmod}{\ensuremath{\mu_{\mathrm{mod}}}}
\newcommand{\mgint}{\ensuremath{\mu_{\mathrm{int}}}}

\newcommand{\Lint}{\ensuremath{L_{\mathrm{int}}}}
\newcommand{\Lobs}{\ensuremath{L_{\mathrm{obs}}}}

\newcommand{\sigL}{\ensuremath{\sigma_L}}
\newcommand{\gamL}{\ensuremath{\gamma_L}}
\newcommand{\xobs}{\ensuremath{X_{\mathrm{obs}}}}
\newcommand{\sigx}{\ensuremath{\sigma_X}}
\newcommand{\xmod}{\ensuremath{X_{\mathrm{mod}}}}
\newcommand{\xint}{\ensuremath{X_{\mathrm{int}}}}
\newcommand{\intx}{\ensuremath{\delta_X}}
\newcommand{\Ctml}{\ensuremath{C_{T\mgas L}}}
\newcommand{\Dtml}{\ensuremath{D_{T\mgas L}}}

\newcommand{\rtm}{\ensuremath{r_{T\mgas}}}
\newcommand{\rtl}{\ensuremath{r_{TL}}}
\newcommand{\rml}{\ensuremath{r_{\mgas L}}}
\newcommand{\mgasm}{\ensuremath{\mgas M}}
\newcommand{\Amm}{\ensuremath{A_{\mgasm}}}
\newcommand{\Bmm}{\ensuremath{B_{\mgasm}}}
\newcommand{\intmm}{\ensuremath{\delta_{\mgasm}}}
\renewcommand{\YM}{\ensuremath{YM}}
\newcommand{\Aym}{\ensuremath{A_{\YM}}}
\newcommand{\Bym}{\ensuremath{B_{\YM}}}

\newcommand{\gamym}{\ensuremath{\gamma_{\YM}}}
\newcommand{\Atm}{\ensuremath{A_{TM}}}
\newcommand{\Btm}{\ensuremath{B_{TM}}}
\newcommand{\gamtm}{\ensuremath{\gamma_{TM}}}
\newcommand{\inttm}{\ensuremath{\delta_{TM}}}
\newcommand{\Alm}{\ensuremath{A_{LM}}}
\newcommand{\Blm}{\ensuremath{B_{LM}}}

\newcommand{\intlm}{\ensuremath{\delta_{LM}}}

\newcommand{\Bqm}{\ensuremath{B_{QM}}}

\begin{document}

\title[Self-consistent cluster scaling relations]{PICACS:
  self-consistent modelling of galaxy cluster scaling relations}
\author[B.J. Maughan]
  {B. J. Maughan\thanks{E-mail: ben.maughan@bristol.ac.uk}\\
  H. H. Wills Physics Laboratory, University of Bristol, Tyndall Ave, Bristol BS8 1TL, UK.
}

\maketitle

\begin{abstract}
  In this paper, we introduce PICACS, a physically-motivated,
  internally consistent model of scaling relations between galaxy
  cluster masses and their observable properties. This model can be
  used to constrain simultaneously the form, scatter (including its
  covariance) and evolution of the scaling relations, as well as the
  masses of the individual clusters. In this framework, scaling
  relations between observables (such as that between X-ray luminosity
  and temperature) are modelled explicitly in terms of the fundamental
  mass-observable scaling relations, and so are fully constrained
  without being fit directly. We apply the PICACS model to two
  observational datasets, and show that it performs as well as
  traditional regression methods for simply measuring individual
  scaling relation parameters, but reveals additional information on
  the processes that shape the relations while providing
  self-consistent mass constraints. Our analysis suggests that the
  observed combination of slopes of the scaling relations can be
  described by a deficit of gas in low-mass clusters that is
  compensated for by elevated gas temperatures, such that the total
  thermal energy of the gas in a cluster of given mass remains close
  to self-similar expectations. This is interpreted as the result of
  AGN feedback removing low entropy gas from low mass systems, while
  heating the remaining gas. We deconstruct the luminosity-temperature
  ($LT$) relation and show that its steepening compared to
  self-similar expectations can be explained solely by this
  combination of gas depletion and heating in low mass systems,
  without any additional contribution from a mass dependence of the
  gas structure. Finally, we demonstrate that a self-consistent
  analysis of the scaling relations leads to an expectation of
  self-similar evolution of the $LT$ relation that is significantly
  weaker than is commonly assumed.
\end{abstract}


\section{Introduction} \label{s.intro}
Simple theoretical arguments
lead to an expectation of power-law scaling relations between the
masses of galaxy clusters and their observable properties
\citep{kai86,bry98}. These scaling relations have been the subject of
a great deal of attention, in particular those involving X-ray
observations of the properties of the intra-cluster medium (ICM)
\citep[e.g.][]{fin01,rei02,san03,vik03}. These X-ray scaling relations
are of interest, as correctly modelling the forms of the scaling
relations tests our understanding of the physical processes that heat
and shape the ICM over cluster lifetimes. Furthermore, if the forms of
the scaling relations are known to some precision, then they provide
an efficient tool to estimate cluster masses in the absence of
detailed data to allow, for instance, an X-ray hydrostatic mass
analysis.

When mass estimates are not available for clusters, it is also common
to study the correlations between X-ray properties as a way to gain
insight into ICM physics. For this reason, the X-ray
luminosity-temperature ($LT$) relation has been extensively studied
\citep[e.g.][]{mit79,edg91,mar98a,pra09a,mau12}. It is widely found
that the slope of the $LT$ relation is steeper than that expected if
gravitational heating of the ICM were the only significant influence
\citep[but see][for a notable exception]{mau12}. Successful models of
additional ICM physics are then expected to explain the steepening of
the $LT$ relation. In this paper, we will argue that fitting models of
the $LT$ relation to observations of clusters and investigating
departures from self-similarity is not optimal. A much better approach
is to jointly model the scaling relations between cluster observables
and their masses, and use those to dictate the form of the $LT$
relation.

The usual approach in these endeavours is to model each of the scaling
relations independently using a form of linear regression. Perhaps the
most popular form is the BCES method, which accounts for errors in the
$x$ and $y$ variables, and intrinsic scatter in the population being
modelled \citep{akr96}. More recently, Bayesian techniques have been
employed to allow greater flexibility in modelling
\citep{kel07,and10}, but these are usually only employed to look at
individual scaling relations. Bayesian techniques are now commonly
used in cosmological studies, with many of these jointly modelling one
or more mass-observable scaling relation
\citep[e.g.][]{roz09,ben11}. The most advanced treatment of the X-ray
scaling relations thus far is the self-consistent modelling of the
scaling relations and mass function for a large sample of clusters by
\citet{man10a,man10b}, which included joint modelling of the mass
scaling of $L$ and $T$.

There is far more information in the cluster datasets than is
typically used in studies of the scaling relations of cluster
populations. In this paper, we present a method of
Physically-motivated, Internally Consistent Analysis of Cluster
Scaling (PICACS) that jointly constrains the form of the scaling
relations between different observables and cluster masses. This
maximises the use of the observational data, provides new information
on the extent to which different observable properties depart from
self-similar behaviour, and gives improved mass estimates for
individual clusters.

The paper is laid out as follows. In \textsection
\ref{sec:self-cons-scal} we derive the set of scaling relations used
to model the cluster population, and then present the statistical
framework used to implement the model in \textsection
\ref{sec:stat-fram}. We then apply the new technique to observed
samples of galaxy clusters with individual hydrostatic mass
measurements (\textsection \ref{sec:mobs}) and without mass estimates
(\textsection \ref{sec:mfit}). We examine the implications of our
results for estimating clusters masses in \textsection
\ref{sec:picacs-mass-estim}, and for dissecting the traditional $TM$
and $LT$ relations in \textsection \ref{sec:picacs-tm-relation} and
\textsection \ref{sec:picacs-lt-relation} respectively. We finish with
a discusion of the limitations of the methodology in \textsection
\ref{sec:cautions-caveats}, before summarising our main results and
conclusions in \textsection \ref{sec:summary-conclusions}. Throughout
the paper, we assume a \LCDM cosmology with $\Omega_M=0.3$,
$\Omega_\Lambda=0.7$ and $H_0=70\kmpspMpc$.

\section{Self-Consistent Scaling Relations}\label{sec:self-cons-scal}
The galaxy cluster X-ray scaling relations were proposed by
\citet{kai86}, based on simple arguments of self-similarity for
clusters dominated by gravity, and their derivations have been
extensively covered in the literature
\citep[e.g.][]{bry98,mau06a}. Here we briefly review the standard
derivations of self-similar scaling relations, and then extend them to
build a self-consistent set of relations to describe non-self-similar
clusters.

The three main properties of the ICM that are observable in X-rays,
and are expected to scale with cluster mass, are the temperature ($T$),
mass ($\mgas$)\footnote{We adopt the unusual notation of $\mgas$ for gas
  mass rather than e.g. \Mgas\ in order to avoid an abundance of
  subscripts.}, and luminosity ($L$) of the ICM. We will consider how
each of these observables are expected to scale with the total mass
($M$) of a cluster. In order to account for the mass dependence of cluster
size, and the evolving background density field from which clusters
collapse, it is convenient to consider properties within an
overdensity radius $R_\Delta$, which encloses a mean density of
$\Delta\rho_c(z)$. The use of the critical density at the redshift of
the cluster as a reference density introduces an expected evolution
into the resulting scaling relations, parameterised through
\begin{align}
E(z) = \sqrt{\Omega_M(1+z)^3 + (1-\Omega_M-\Lambda)(1+z)^2 + \Lambda}
\end{align}
and which describes the redshift dependence of the Hubble
parameter. This leads to
\begin{align}
R_\Delta & \propto E(z)^{-2/3}M_\Delta^{1/3}
\end{align}
which can be used to eliminate $R$ in favour of $M$.  In the
following derivations, all properties are implicitly measured within
the same radius $R_\Delta$, and we drop the $\Delta$ subscript
  for compactness.

\subsection{The $\mgas M $ Relation}
For self-similar clusters, the mass of gas (\mgas) in the ICM is a
constant fraction \fgas\ of the total mass:
\begin{align}\label{eq:fgm1}
\mgas & = \fgas M
\end{align}
This can be rewritten, with the addition of a slope parameter to allow
a mass dependency, as
\begin{align}\label{eq:mgm}
\frac{\mgas}{\mgas_0} & = \Amm\left(\frac{M}{M_0}\right)^{\Bmm}
\end{align}
Here $A_{\mgas}$ is a constant of proportionality and $\mgas_0$ and $M_0$ are
normalisation constants introduced for numerical convenience
later. Throughout this work, we use $\mgas_0=5\times10^{13}\Msol$ and
$M_0=5\times10^{14}\Msol$.

We have assumed that the gas fraction is constant with redshift.  In
principal an evolution term could be included in equation
\eqref{eq:mgm}, but we defer investigation of the evolution of scaling
relations in the PICACS model to a later paper. As an aside, we note
that equation \eqref{eq:mgm} is equivalent to writing the mass
dependency of \fgas\ as
\begin{align}\label{eq:fgm2}
\fgas & = \frac{\mgas_0}{M_0}\Amm \left(\frac{M}{M_0}\right)^{\Bmm-1}
\end{align}

\subsection{The $TM$ relation}
Assuming that the ICM is in virial equilibrium with the cluster
gravitational potential, the virial theorem gives
\begin{align}
T & \propto \frac{M}{R},
\end{align}
Eliminating $R$ yields the well-known, self-similar $TM$ relation:
\begin{align}\label{eq:TMSS}
T \propto E(z)^{2/3}M^{2/3}.
\end{align}
which we will generalise to give
\begin{align}\label{eq:TM}
\frac{T}{T_0} & = \Atm E(z)^{\gamtm}\left(\frac{M}{M_0}\right)^{\Btm}
\end{align}
Throughout this work, we use $T_0=5\keV$.

\subsection{The $LM$ Relation}
The luminosity of the ICM is dominated by bremsstrahlung emission for
$T\gtrsim2\keV$, where line emission is not significant. If we also
neglect the weak temperature dependence of the Gaunt factor, then the
luminosity is given by
\begin{align}\label{eq:brem}
L & \propto \int \rhog^2 T^{1/2} dV
\end{align}
The integral of the gas density, \rhog, depends on the distribution of
the ICM. We follow \citet{arn99} by factoring the density into a mean
density and a dimensionless structural parameter
$Q=\left<\rhog^2\right>/\left<\rhog\right>^2$ (where angle
  brackets indicate volume averages), such that equation
(\ref{eq:brem}) becomes
\begin{align}\label{eq:LM}
L & \propto E(z)^2 Q\fgas^2 T^{1/2} M
\end{align}
In other words, the luminosity of the ICM depends both on the amount
of gas in the cluster (via \fgas), and how that gas is distributed
(via $Q$). For self-similar clusters, $Q$ and
\fgas\ are independent of mass and can be absorbed into the
proportionality constant.

It is widespread practice \citep[e.g.][]{mau06a} to derive the
self-similar relation between luminosity and mass ($LM$ relation) by
setting $Q$ and \fgas\ to constants and using equation (\ref{eq:TMSS})
to eliminate $T$ from equation (\ref{eq:LM}). However, this does not
maximise the observational information, and if self-similar behaviour
breaks down, it becomes unclear which of the mass scalings are being
broken.

Instead, we rewrite equation (\ref{eq:LM}) to allow for a power-law
mass dependence of the ICM structure parameter $Q$ (moving any
constant component into the proportionality constant), giving
\begin{align}
L & \propto E(z)^2 \fgas^2 T^{1/2} M^{\Bqm}
\end{align}
Although we interpret this slope parameter $\Bqm$ as predominantly
describing mass-dependence of $Q$, it could also describe the effects
of the increasing contribution of line emission to the luminosity at
lower temperatures ($\lesssim 2\keV$), which modifies the temperature
dependence towards $T^{-1/2}$ in equation
\eqref{eq:brem}.

Now, rather than use equation (\ref{eq:fgm1}) to substitute for \fgas,
let us instead explicitly keep the observed quantities, and write
the bremsstrahlung relation as
\begin{align}\label{eq:ltm}
  \frac{L}{L_0} & = A_L
  E(z)^{\gamma_L}\left(\frac{\mgas}{\mgas_0}\right)^{2}
  \left(\frac{T}{T_0}\right)^{1/2} \left(\frac{M}{M_0}\right)^{\Bqm-2}
\end{align}
where $\gamma_L=2$ and $\Bqm=1$ for self-similar clusters, and we set
$L_0=5\times10^{44}\ergps$.

The traditional $LM$ relation is then given by substituting equations
(\ref{eq:mgm}) and (\ref{eq:TMSS}) to eliminate $\mgas$ and $T$ in
favour of $M$:
\begin{align}\label{eq:ltm2}
  \frac{L}{L_0} & = A_{LM}E(z)^{\gamma_{LM}} \left(\frac{M}{M_0}\right)^{B_{LM}}
\end{align}
where
\begin{align}
A_{LM} & = A_L \Amm^2\Atm^{1/2} \label{eq:alm}\\
B_{LM} & = 2\Bmm + \frac{1}{2}\Btm + \Bqm -2 \label{eq:blm}\\
\gamma_{LM} & = \gamL + \frac{1}{2}\gamtm
\end{align}
For self-similar clusters in virial equilibrium, $B_{LM}=4/3$ and $\gamma_{LM}=7/3$.

Other, composite, X-ray scaling relations may then be produced by combining the preceding scaling relations, as in the following subsections.

\subsection{The $Y_XM$ Relation}
The product of \mgas\ and $T$ is proportional to the total thermal
  energy content of the ICM and is usually termed $Y_X$, which has
  been shown to follow a low-scatter correlation with mass
  \citep[e.g.][hereafter A07]{kra06a,mau07b,arn07}. The $Y_XM$
  relation is obtained by combining the $\mgas M$ and $TM$ relations:
\begin{align}\label{eq:ym}
\frac{Y_X}{Y_{X0}} & = \Aym E(z)^{\gamym}\left(\frac{M}{M_0}\right)^{\Bym}
\end{align}
where $Y_{X0}=2.5\times10^{14}\Msol\keV$ and
\begin{align}
\Aym & = \Atm\Amm \\
\Bym & = \Btm - \Bmm \\
\gamym & = \gamtm
\end{align}
For self-similar clusters in virial equilibrium, $\gamym=2/3$ and
$\Bym=5/3$. A07 argued that the $Y_X$ may be the ICM property most
closely related to the cluster mass, in which case any deficit of gas
in the cluster potential (due to its removal or incomplete accretion)
would be balanced by an increase in temperature to leave the total
thermal energy unchanged. In this case, a difference in $\Bmm$ from
unity would be compensated for by a corresponding change in $B_{TM}$
to maintain $\Bmm+B_{TM}=\Bym=5/3$.

\subsection{The $LT$ Relation}
Finally, the relation between luminosity and temperature ($LT$
relation) has long been used as a key observational diagnostic of
non-gravitational processes in clusters, with departures from the
self-similar form of the $LT$ relation used to measure the nature and
extent of those processes. However, as for the $LM$ relation, the
self-similar form of the $LT$ relation is usually derived by
substituting equation (\ref{eq:TMSS}) into equation (\ref{eq:LM}) to
eliminate $M$, and then assuming $Q$ and \fgas\ are constant with
  mass. This results in an $LT$ relation of the form
\begin{align}\label{eq:lt}
\frac{L}{L_0} & = A_{LT}E(z)^{\gamma_{LT}}\left(\frac{T}{T_0}\right)^{B_{LT}}
\end{align}
where departures from $\gamma_{LT}=1$ and $B_{LT}=2$ are taken as
evidence for similarity breaking. However, this is only true if all of
the scaling relations between the cluster observables and mass are
self similar. This is made clear if we write the parameters of the
$LT$ relation in terms of the PICACS scaling relations:
\begin{align}
A_{LT} & = A_L \Amm^2 \Atm^{1/2 - (2\Bmm+\Btm/2+\Bqm-2)/\Btm} \label{eq:alt}\\
 & = A_{LM}A_{TM}^{-B_{LM}/B_{TM}} \nonumber \\
\nonumber \\
B_{LT} & = \frac{2\Bmm + \Btm/2 + \Bqm - 2}{\Btm} \label{eq:blt}\\
 & = B_{LM}/B_{TM} \nonumber \\
\nonumber \\
\gamma_{LT} & = \gamma_L + \gamtm\left(\frac{1}{2} -
  \frac{2\Bmm + \Btm/2 + \Bqm -
    2}{\Btm}\right) \label{eq:glt} \\
 & = \gamma_{LM} - \frac{B_{LM}}{B_{TM}}\gamma_{TM} \nonumber
\end{align}
Thus the slope of the $LT$ relation departs from self-similarity if
any or all of the slopes of the fundamental scaling relations differ from
their self-similar values, but measuring the slope of the $LT$
relation will not tell us which. Similarly, the self-similar evolution
of the $LT$ relation differs from $\gamma_{LT}=1$ if the evolution
{\em or} slopes of the fundamental scaling relations differ from their
self-similar values. In other words, a simple measurement of the
evolution of the $LT$ relation could imply real evolution, when in
fact the fundamental scaling relations evolved self-similarly, but the
slope of one or more were not self-similar.

\subsection{The PICACS Scaling Relations}
Equations (\ref{eq:mgm}), (\ref{eq:TM}) and (\ref{eq:ltm}) form a
physically-motivated, internally consistent description of the
fundamental scaling relations between the key X-ray observables and
cluster mass. The composite scaling relations in equations
(\ref{eq:ltm2}) and (\ref{eq:lt}) are also well-established, but the
explicit dependencies on the fundamental scaling relations are not
usually preserved, losing information as a result. We refer to these
composite relations, with those dependencies explicitly preserved, as
the PICACS scaling relations.

In this paper, We argue that the traditional modelling of the
  $\mgas M$, $LT$, $LM$, and $TM$ relations without recognising their
  dependencies on the fundamental mass scaling relations is a tool
that is at best blunt, but possibly also inaccurate, for the study of
cluster scaling relations. Instead we propose the use of the
  PICACS approach, by which we refer to the joint modelling of cluster
  populations with the PICACS scaling relations. In the following
  section we present a statistical framework to enable this
  modelling.

\section{Statistical Framework}\label{sec:stat-fram}
Perhaps the most obvious way to measure the PICACS scaling relations
would be to fit each relation independently to a sample for which we
have observations of $T$, \mgas, $L$, and an observationally
determined $M$ (e.g. from X-ray hydrostatic masses). However, by
using Bayesian techniques, it is possible to construct a statistical
framework to jointly determine the probability distributions of the PICACS
parameters and cluster masses, given the observational data. The
following treatment was inspired by the Bayesian analysis of cluster
mass-richness relations in \citet{and10}.

Generically, Bayes' theorem can be used express the probability of some model
parameters $\theta_i$ given observational data $D_j$ as
\begin{align}
P(\theta_i|D_j) & \propto P(D_j|\theta_i)P(\theta_i)
\end{align}
The probability on the left hand side is referred to as the {\em
  posterior}, while the first term on the right describes the {\em
  likelihood} (\lik) of the data given the model. The last term describes the
{\em prior} probabilities of the model parameters.

We can construct the likelihood of the PICACS scaling relations and
cluster masses in terms of the observables. The PICACS scaling
relations predict the value of each observable given a cluster mass,
but observed values are expected to differ from the model predictions
due to the intrinsic scatter $\inttm$, $\intmm$, $\intlm$ of the
population about each relation, and the statistical scatter described
by the measurement errors $\sigt$, $\sigmg$, $\sigL$ on
each observed quantity. In the following, we will use the subscripts
$obs$ to indicate an observed quantity, $mod$ to indicate a quantity
predicted by a PICACS scaling relation, and $int$ to indicate the
model prediction including intrinsic scatter.

For example, for a cluster of mass $M$, we might have an observed mass
$M_{obs}$ with error $\sigm$, and a predicted temperature
$T_{mod}$ from equation (\ref{eq:TM}). The intrinsic scatter in the
relation $\inttm$ will then randomly shift the temperature to a
value $T_{int}$, which we then observe as $T_{obs}$ with error
$\sigt$.

The likelihood of our observation of $M_{obs}$ for a cluster of mass
$M$ is simply given by
\begin{align}
\lik_M & = P(\mobs|M,\sigm)
\end{align}

The likelihood of our observation of \tobs\ for the same cluster is the
product of the probabilities of the cluster being scattered to
temperature $\tint$ and then observed at temperature $\tobs$:
\begin{align}
\lik_T & = P(\tobs|\tint,\sigt)P(\tint|\tmod,\inttm)  \nonumber \\
       & = P(\tobs|\tint,\sigt)P(\tint|M,\theta_T,\inttm)
\end{align}
where \tmod\ is the temperature predicted by the $TM$ scaling relation in
Equation (\ref{eq:TM}), so is a function of $M$ and the scaling
relation parameters $\theta_T=(A_{TM}, B_{TM}, \gamma_{TM})$.

The likelihood of the observation of the gas mass \mgas\ for the same
cluster is similarly
\begin{align}
\lik_\mgas & = P(\mgobs|\mgint,\sigmg)P(\mgint|\mgmod,\intmm)
\nonumber \\
       & = P(\mgobs|\mgint,\sigmg)P(\mgint|M,\theta_\mgas,\intmm)
\end{align}
where \mgmod\ is given by equation (\ref{eq:mgm}), and is a function
of $M$ and the scaling relation parameters $\theta_\mgas=(\Amm, \Bmm)$.

Finally, the likelihood function for $L$ is
\begin{align}
\lik_L & = P(\Lobs|\Lint,\sigL)P(\Lint|L_{mod},\intlm) \nonumber \\
 & = P(\Lobs|\Lint,\sigL)P(\Lint|M,\theta_L,\intlm)
\end{align}
where $\theta_L=(A_{LM}, B_{LM},\gamma_{LM})$ are the PICACS $LM$ scaling relation parameters.

\subsection{Modelling Covariance}
The likelihood expressions derived above assume that the
  intrinsic scatter and statistical scatter on the observables are all
  independent The PICACS scaling relations make clear that the
  intrinsic scatter terms should not be independent. For example,
  intrinsic scatter in $\mgas$ at a given $M$ will contribute to
  scatter in both the $TM$ and $LM$ relations. Furthermore, the
  processes driving the intrinsic scatter (e.g. mergers, cooling, AGN
  feedback) will impact all of the ICM observables to a greater or
  lesser extent.

  The covariance of intrinsic scatter has not been
  widely studied, but simulations suggest a coherent motion of
  clusters along the $LT$ relation during mergers
  \citep[e.g.][]{row04,har08}, implying correlated scatter in the $TM$
  and $LM$ relations. The covariance of several cluster observables
  has also been investigated in the simulations of \citet{sta10} and
  \citet{ang12}. Observationally, \citet[][]{man10b} found the
  correlation of intrinsic scatter in $L$ and $T$ to be consistent
  with zero, albeit without strong constraints.

  The possibility of correlated intrinsic scatter is incorporated into
  the PICACS model by using a covariance matrix \Ctml\ to describe the
  intrinsic scatter. The diagonal terms are
  $\inttm^2,\intmm^2,\intlm^2$, while the off-diagonal terms give the
  covariances between $\tint,\mgint$ and $\Lint$. The joint likelihood
  of the intrinsically scattered values is now
\begin{align}
\lik_{int} = P(\tint,\mgint,\Lint|M,\theta_L,\theta_T,\theta_\mgas,\Ctml)
\end{align}
where the probability distribution is a multivariate Gaussian
distribution with mean values given by $(\tmod,\mgmod,L_{mod})$ and
covariance given by \Ctml. With this change, the full covariance
matrix becomes a parameter of the model.

Here we have explicitly assumed that the intrinsic scatter in cluster
properties is log-normal in form. This is supported by the results of
\citet{mau07b}, who found that the intrinsic scatter in core-excised
luminosities (as used in the present study) is consistent with a
log-normal distribution, and \citet{vik09} who found that the
intrinsic scatter of {\em core-included} luminosities is also
consistent with a log-normal distribution. However, it has been found
that cool-core related properties of clusters show a bimodal
distribution \citep{san09}, suggesting that a log-normal distribution
for the intrinsic scatter of core-included properties is an imperfect
(though reasonable) assumption.

\begin{table*}
\begin{center}
\begin{tabular}{lcccccl}
  \hline
Name & z & $T$ & \mgas & \mobs & \mfit & Reference \\
       &   & keV & $10^{13}\Msol$ & $10^{14}\Msol$ & $10^{14}\Msol$ \\
  \hline
A133        & 0.0569 & $4.02\pm0.07$ & $2.82\pm0.34$   & $3.26\pm0.39$   &  $3.06 \pm 0.25$  & V06 \\
A383        & 0.1883 & $4.67\pm0.12$ & $4.07\pm0.41$   & $3.15\pm0.32$   &  $3.42 \pm 0.27$  & V06 \\
A478        & 0.0881 & $7.70\pm0.12$ & $9.89\pm1.30$   & $7.90\pm1.04$   &  $8.19 \pm 0.74$  & V06 \\
A1413       & 0.1429 & $7.16\pm0.11$ & $8.69\pm0.87$   & $7.79\pm0.78$   &  $7.51 \pm 0.58$  & V06 \\
A1795       & 0.0622 & $5.94\pm0.05$ & $6.73\pm0.58$   & $6.20\pm0.53$   &  $5.97 \pm 0.40$  & V06 \\
A1991       & 0.0592 & $2.53\pm0.06$ & $1.35\pm0.19$   & $1.27\pm0.17$   &  $1.37 \pm 0.13$  & V06 \\
A2029       & 0.0779 & $8.22\pm0.09$ & $10.57\pm0.98$  & $8.24\pm0.76$   &  $8.61 \pm 0.66$  & V06 \\
A2390       & 0.2302 & $8.62\pm0.17$ & $16.25\pm1.63$  & $11.05\pm1.11$  &  $10.82 \pm 0.89$ & V06 \\
MKW4        & 0.0199 & $1.59\pm0.04$ & $0.51\pm0.07$   & $0.79\pm0.10$   &  $0.72 \pm 0.07$  & V06 \\
A1983       & 0.0442 & $2.08\pm0.09$ & $0.64\pm0.09$   & $1.09\pm0.37$   &  $0.91 \pm 0.13$  & A07    \\
MKW9        & 0.0382 & $2.32\pm0.23$ & $0.49\pm0.05$   & $0.88\pm0.20$   &  $0.80 \pm 0.12$  & A07    \\
A2717       & 0.0498 & $2.44\pm0.06$ & $1.02\pm0.05$   & $1.10\pm0.12$   &  $1.18 \pm 0.10$  & A07    \\
A2597       & 0.0852 & $3.50\pm0.09$ & $2.51\pm0.09$   & $2.22\pm0.22$   &  $2.34 \pm 0.19$  & A07    \\
A1068       & 0.1375 & $4.46\pm0.11$ & $3.77\pm0.10$   & $3.87\pm0.28$   &  $3.66 \pm 0.23$  & A07    \\
PKS0745-191 & 0.1028 & $7.61\pm0.27$ & $10.71\pm0.48$  & $7.27\pm0.75$   &  $7.75 \pm 0.63$  & A07    \\
A2204       & 0.1523 & $7.89\pm0.21$ & $10.55\pm0.40$  & $8.39\pm0.81$   &  $8.25 \pm 0.61$  & A07    \\
   \hline
\end{tabular}
\caption{\label{tab:vamfit}X-ray properties of the VA sample clusters
  taken from V06 and A07 with the addition of the cluster masses
  (\mfit) determined with PICACS (\mobs\ are the hydrostatic from the
  referenced works). The V06 values have been rescaled from
    $H_0=72\kmpspMpc$ to $H_0=70\kmpspMpc$. Where the uncertainties
  on the A07 properties were asymmetric, the mean value is used. $T$
  values have been rescaled to the $[0.15-1]\rf$ aperture as described
  in the text. The $T$ and \mgas\ values have not been scaled by
    the cross calibration factors introduced in the text.}
\end{center}
\end{table*}

It is also possible to model the effect of covariance in the
statistical scatter. This covariance matrices would ideally be known
from the analysis of the data, but for the literature data used for
the current study, these were not available, and so statistical errors
were treated as being independent. In principal it is possible to
include the covariance in statistical scatter as additional free
parameters in the model, and for completeness we present a strategy
for doing so below. However, for the data analysed here, this
additional complexity was found to be computationally expensive while
making no significant change to the model fits, and was thus
neglected.

Describing the statistical scatter with a single covariance matrix is
not possible, as the matrix would be different for each cluster
due to the differing statistical errors. Instead, the covariance of
the statistical scatter can be modelled in terms of the
correlation coefficients between the statistical errors on the
observed quantities, \rtm, \rtl, \rml. The joint likelihood of the
observed values of each property would then given by
\begin{align}
\lik_{obs} = P(\tobs,\mgobs,\Lobs|\tint,\mgint,\Lint,\Dtml)
\end{align}
where the probability distribution is a multivariate Gaussian with
mean $\tint,\mgint,\Lint$ and covariance given by \Dtml, which is
defined for each cluster with diagonal elements
$\sigt^2,\sigmg^2,\sigL^2$, and off-diagonal elements
$\rtm/(\sigt\sigmg), \rtl/(\sigt\sigL), \rml/(\sigmg\sigL)$. Here the
$\sigma$ terms are different for each cluster (the measurement
errors), but the correlation coefficients $r$ are in common. The three
correlation coefficients are thus the model parameters describing the
covariance of statistical scatter.

Neglecting the covariance in the statistical scatter, the
  likelihood of the observed values is just
\begin{align}
\lik_{obs} = P(\tobs|\tint,\sigt)P(\mgobs|\mgint,\sigmg)P(\Lobs|\Lint,\sigL)
\end{align}

\subsection{The Final Joint Likelihood}
For a set of observations of multiple clusters (denoted by the index
$i$), we take the product of each likelihood evaluated over all
clusters. The joint likelihood of our observations is then given by
\begin{align}
\lik & = \prod_i \lik_{M,i} \lik_{int,i} \lik_{obs,i}
\end{align}

The posterior probability distribution of our model parameters is
\begin{multline}\label{eq:post}
P(\theta_T,\theta_\mgas,\theta_L,M,\tint,\mgint,\Lint|\mobs,\tobs,\mgobs,\Lobs)
\\
\propto \lik\, P(\theta_T)P(\theta_\mgas)P(\theta_L)P(M)P(\Ctml)
\end{multline}

Note that the cluster mass $M$ is a parameter of our model, as are the
intrinsically scattered quantities. The latter are simply nuisance
parameters that are marginalised over, but the appearance of $M$ as a
parameter means that the PICACS framework can be used to constrain $M$
given a combination of $\mobs$ and/or priors on the form of some or
all of the scaling relations. A key advantage of this method is that
all of the observables are being fit against the same cluster mass in
an internally consistent manner.

As we will demonstrate later, it is entirely possible to fit the
PICACS relations to the observed quantities without an observed mass
for the clusters being considered. In this case, we remove $\lik_M$
from our likelihood function, but keep $M$ as a parameter of the
model. $M$ can then be constrained by the PICACS scaling relations,
but care is needed as strong degeneracies arise between the scaling relation
parameters and between the intrinsic scatter terms. To avoid this,
informative priors are required on a subset of the parameters, as
discussed in \textsection \ref{sec:covar-degen-1}.

PICACS shares some features and much of its philosophy with the
\citet{man10b} approach, but is not designed with cosmological
analyses in mind, and so currently lacks the ability to model
selection biases, which is a major strength of the \citet{man10b}
work. On the other hand, PICACS has a more detailed model of the
inter-dependency of the scaling relations (specifically the inclusion
of a variable slope in the $\mgas M$ relation and the explicit
inclusion of \mgas\ in the $TM$ relations, and $T$ and \mgas\ in
the $LM$ relation).

\subsection{Implementation}
In order to obtain constraints on our model parameters, the posterior
distribution must be sampled over the large parameter space. There are
many tools available for this generic problem, and the PICACS
framework could be implemented in many ways. Here we note a few of the
specifics of our implementation. PICACS was developed using the {\em
  R} statistical computing environment \citep{r12}, and the posterior
probability distribution was analysed using the Bayesian inference
package {\em Laplace's Demon} within
$R$\footnote{\url{http://www.bayesian-inference.com/software}}. {\em
  Laplace's Demon} contains many Markov Chain Monte Carlo (MCMC)
algorithms designed to efficiently sample the posterior probability
distribution, and the ``Adaptive Metropolis-within-Gibbs'' algorithm
was found to be effective at sampling the PICACS posterior
distribution and converging reliably. We refer the reader to the
excellent documentation in Laplace's Demon for details of the algorithm,
as well as an introduction to Bayesian inference.

It is computationally advantageous to rewrite the PICACS scaling
relations in log space (we used log$_{10}$ for convenience). Thus each
probability distribution in the likelihood functions is implemented as
a Gaussian distribution in log$_{10}$ space. It is also essential to
work with the logarithm of all probabilities, due to the small
numerical values involved, and hence we sampled the natural logarithm
of the posterior probability distribution, and all of the products in
the likelihood and posterior expressions become sums. Unless otherwise
stated, a flat prior was assumed for all parameters. Three MCMC chains
were run in parallel with randomised initial values and the fits were
accepted when the three chains had converged (as determined by
examination of the sample distributions). The probability distribution
of each parameter was computed from the distribution of samples from
the chain after removing the start of each chain until the parameter
values were stationary, and combining multiple chains. We summarise
the posterior probability distribution of each parameter as the mean,
plus or minus the standard deviation of each distribution.

The presence of covariance matrices in the model significantly
increases the computational load of evaluating the likelihood. This is
because each proposed covariance matrix must be tested and rejected if
it is not positive definite. This is particularly burdensome for the
case of covariance in the statistical scatter, where the proposed
correlation coefficients give a different covariance matrix for each
cluster, due to the differing measurement errors for each cluster.

In the following sections we apply the PICACS framework to
observational samples to demonstrate its applicability and
effectiveness in several scenarios.

\section{Application to clusters with observed masses}\label{sec:mobs}
As a first test, we apply the PICACS framework to a set of clusters
with precise mass estimates from X-ray hydrostatic analyses, and
compare the performance to that of traditional BCES regression
fits. For this study we used the samples of \citet[][hereafter
V06]{vik06a} and A07. Both samples target
relaxed, low-z clusters with high quality data over a reasonably large
range in mass, so are well suited to a first test of PICACS. Both
samples have published $\mobs$, $\tobs$ and $\mgobs$ for each cluster,
but neither sample has published luminosities available. We thus
removed $\lik_L$ and $\theta_L$ from PICACS, leaving $\theta_T$,
$\theta_\mgas$, $C_{T \mgas}$ (the covariance between \tint\ and
\mgint), \rtm\ (the correlation between statistical scatter in $T$ and
\mgas) and $M$ as the parameters of interest.

\begin{figure*}
\begin{center}
\scalebox{0.43}{\includegraphics*{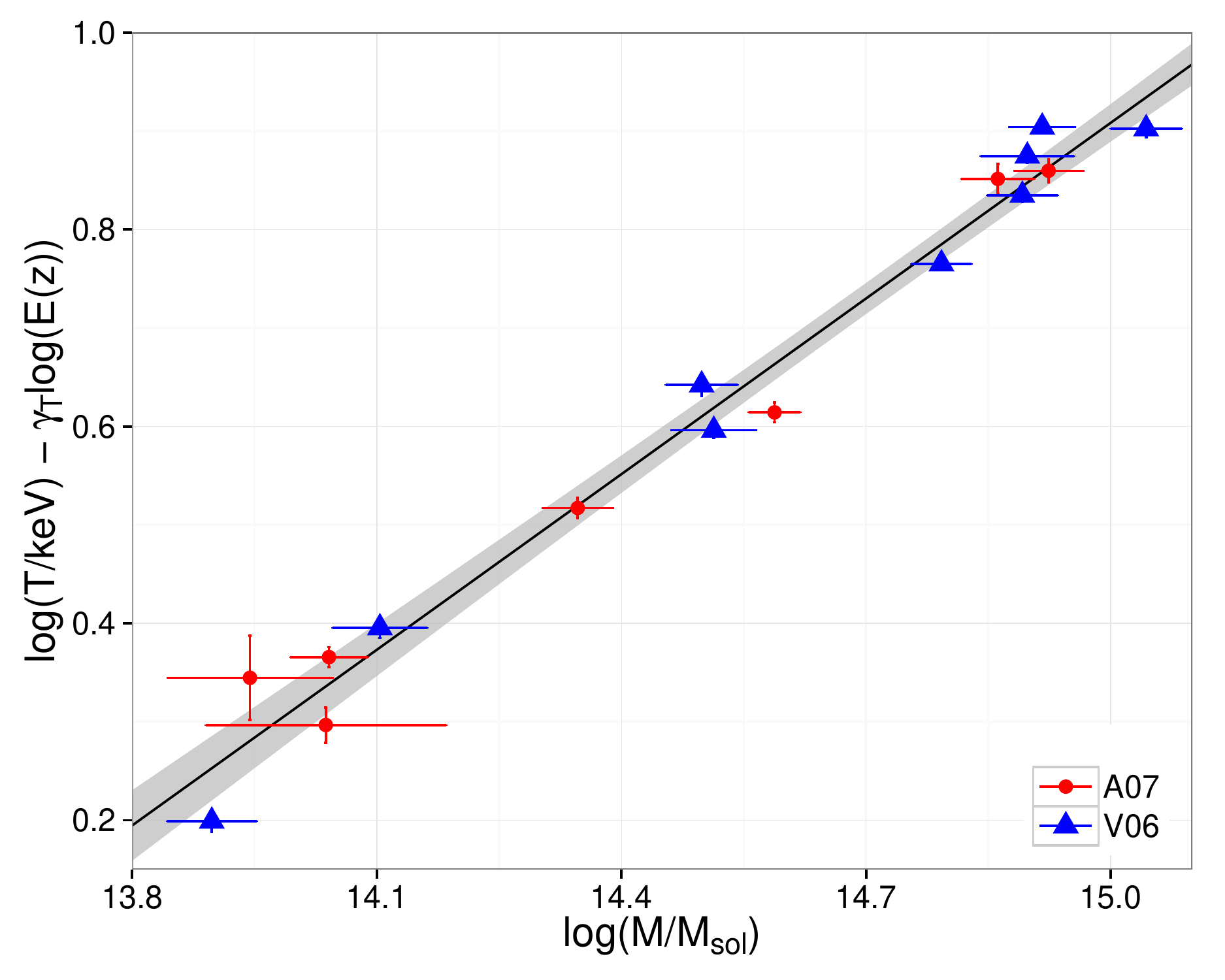}}
\hspace{0.5cm}
\scalebox{0.43}{\includegraphics*{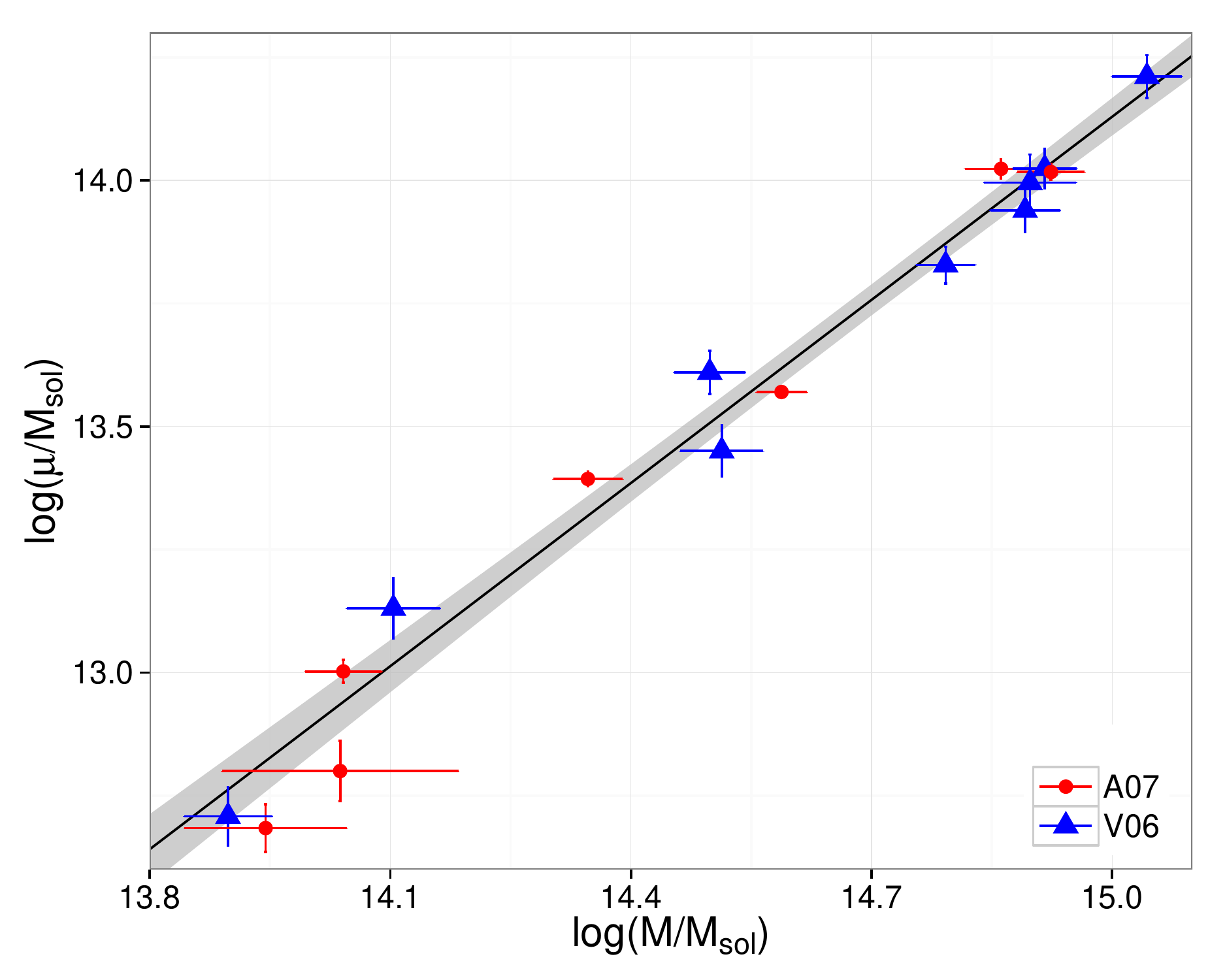}}
\caption[]{\label{f.va} Best fitting PICACS fits to the $TM$ (left)
  and $\mgas M$ (right) scaling relations of the VA sample, with the
  shaded envelope indicating the $1\sigma$ uncertainty. The A07
  clusters have been scaled by the best-fitting $F_T$ and $F_\mgas$
  cross-calibration parameters. Note that in these plots, the mass of each
  cluster is \mfit, the mass determined by the combined PICACS fit of
  the masses and scaling relations.}
\end{center}
\end{figure*}

\begin{table*}
\begin{center}
\scalebox{0.85}{
\begin{tabular}{lcccccccccc}
  \hline
  Method & $\Atm$ & $\Btm$ & \inttm & $F_T$ & $\Amm$ & $\Bmm$ & \intmm
  & $F_\mgas$ & $\Bym$ & $\rho_{T\mgas}$\\
  \hline
  PICACS & $1.07\pm0.04$ & $0.59\pm0.04$ & $0.08\pm0.03$ & $0.97\pm0.06$
  & $1.14\pm0.08$ & $1.24\pm0.07$ & $0.14\pm0.06$ & $0.99\pm0.11$ &
  $1.84\pm0.10$ & $0.3\pm0.5$ \\

  BCES & $1.10\pm0.02$ & $0.60\pm0.03$ & $0.01\pm0.01$ & - &
  $1.15\pm0.04$ & $1.29\pm0.05$ & $0.02\pm0.04$ & - & - & - \\
  \hline
\end{tabular}
}
\caption{\label{tab:mobs} Best fitting parameters of the $TM$ and
  $\mgas M$ relations fit to the VA sample using the PICACS and
  orthogonal BCES methods. For convenience, the intrinsic scatter
  terms are given in natural log space, so are simply fractional
  values. For the PICACS fit, the intrinsic scatter values are
  computed from the diagonal elements of the covariance matrix. $F_T$
  and $F_\mgas$ are factors introduced to model relative calibration
  errors between the instruments used (see text). The final column
  $\rho_{T\mgas}$, gives the Pearson's correlation for the parameters,
  derived from the covariance matrix.}
\end{center}
\end{table*}

The two samples were combined, and duplicate clusters were removed
from the A07 sample, and three clusters without measurements at \rf\
were excluded from the V06 sample. In addition, Abell 907 was removed
from the combined list, as it appears in the REXCESS sample
\citep{boh07}, for which we will be using the constraints from this
study as independent priors in our subsequent analysis. This gave a
combined list of 16 clusters at $0.02<z<0.23$ with a median $z=0.09$,
which we refer to as the ``VA sample''. The properties of the sample
used for this study are summarised in Table \ref{tab:vamfit}.

The temperatures from A07 were measured in the $[0.15-0.75]\rf$
aperture so were scaled to the $[0.15-1]\rf$ aperture by multiplying
them by $0.955$, the midpoint of the range suggested by A07. Similarly
the V06 temperatures were rescaled from the $70\kpc - \rf$ aperture in
which they were measured, to the $[0.15-1]\rf$ aperture by multiplying
them by $0.97$ as recommended in V06. The gas and total masses were
all measured within \rf.

As the two parent samples were observed with different satellites,
\Chandra\ (V06) and \XMM\ (A07), we introduce additional
cross-calibration factors, $F_T$ and $F_\mgas$, such that the A07
\tobs\ and \mgobs\ were multiplied by these factors respectively. We
note that the choice to rescale the A07 rather than V06 properties was
arbitrary, but as we shall see, these factors turn out to be
negligible. The ability to include additional model components such as
these scale factors is an advantage of the PICACS approach over BCES
regression.

The PICACS models were fit to the VA sample, and the resulting
$TM$ and $\mgas M$ scaling relations are plotted in
Figure \ref{f.va}. The fits were also performed with the standard
orthogonal BCES method, and with PICACS assuming independent scatter
in the observables. The model parameters obtained with these
techniques are summarised in Table \ref{tab:mobs}. In these fits, the
evolution parameters $\gamtm$ and $\gamma_L$ were fixed at their
self-similar values due to the small redshift range of the sample.

The PICACS fits agree extremely well with the results from the
conventional BCES fits (with the exception of the scatter
  measurements, discussed in the following section), demonstrating
that the new technique performs well.

\begin{figure*}
\begin{center}
\scalebox{1.0}{\includegraphics*{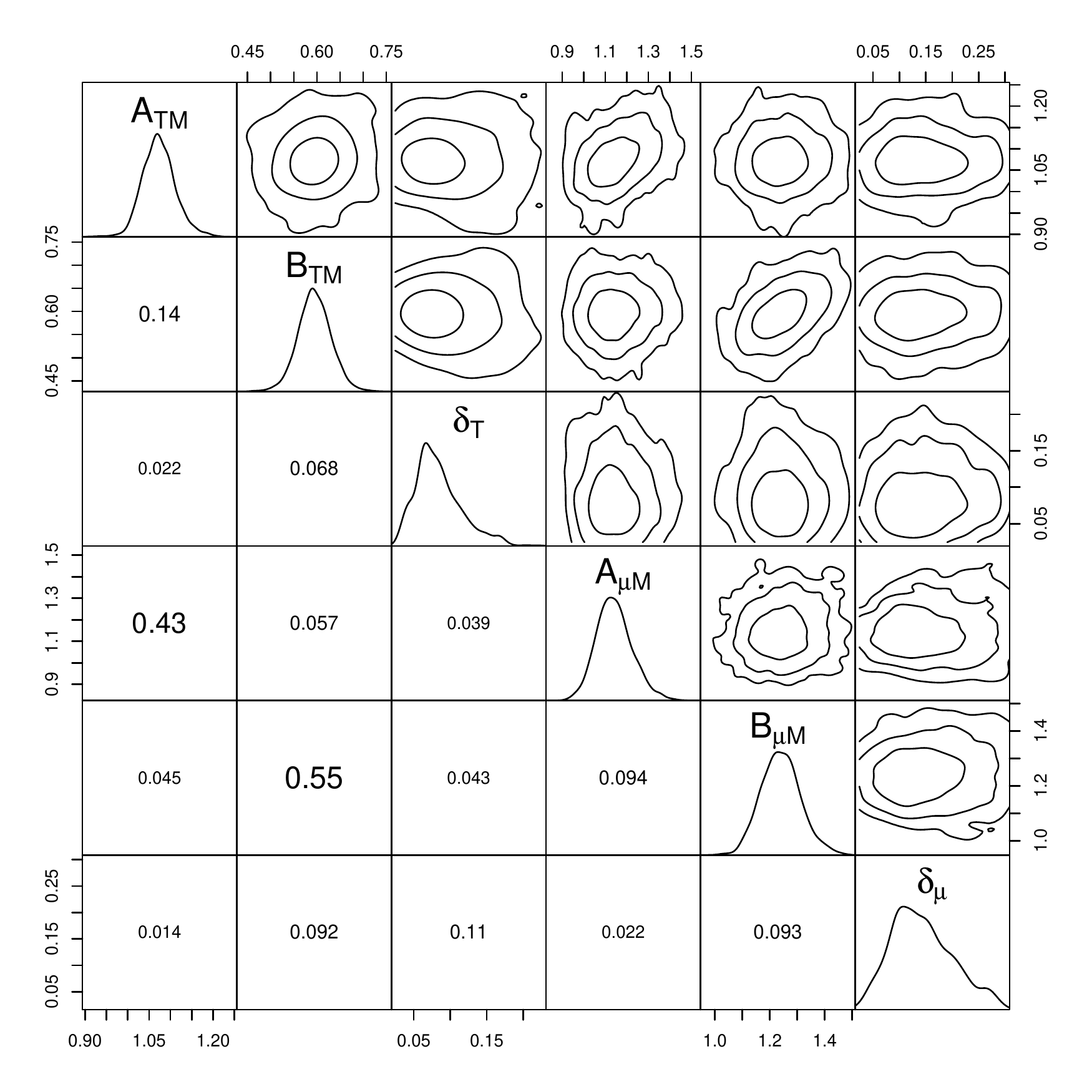}}
\caption[]{\label{f.vamat} Correlation matrix of the PICACS model
  parameters for the fit to the VA sample. The posterior densities are
  shown along the diagonal, with $1\sigma$, $2\sigma$, and
    $3\sigma$ confidence contours for the pairs of parameters shown on
    the upper triangle panels. The lower triangle panels show the
  Pearson's correlation coefficient for the corresponding pair of
  parameters (with a text size proportional to the correlation
  strength). The $\delta$ terms are in natural log space, and were
  computed from the square root of the diagonal elements of the
  covariance matrix $C_{T_\mgas}$ and so do not represent the full
  information in the covariance matrix.}
\end{center}
\end{figure*}

\subsection{Covariance and Degeneracies}\label{sec:covar-degen}
Table \ref{tab:mobs} shows that the measurements of the intrinsic
scatter in the $TM$ and $\mgas M$ relations differ significantly
between the PICACS and BCES methods. However, the definitions of
intrinsic scatter also differ, in that the BCES method does not
measure scatter itself. Instead, we follow \citet{mau07b} in defining
the intrinsic scatter as the constant term that must to be added to
the combined $T$ (or \mgas) and $\mobs$ error bars in quadrature
in log$_{10}$ space to produce a reduced $\chisq$ of unity in the
  $T$ or \mgas\ direction with respect to the BCES regression
line. This is not self-consistent, as the best fitting BCES model is
not the model which minimises the $\chisq$ in any
direction. Furthermore, this approach treats the intrinsic scatter in
each relation independently.

In the case of PICACS, if a single scaling relation were fit, then the
model masses would move to minimise the scatter in the relation,
subject to \mobs\ and its error. In this case, the scatter measured
was found to be consistent with the BCES value. However, when PICACS
is jointly fitting $TM$ and $\mgas M$ relations, the model masses must
satisfy both relations, and could only reduce the scatter
significantly below the raw scatter in the data if there were strong
positive covariance in the scatter in $T$ and \mgas. In other words,
the model masses from PICACS are not the masses which minimise the
scatter in either the $TM$ or $\mgas M$ relation alone; they are the
masses which jointly satisfy both relations in a consistent way. Using
either relation alone would give a lower scatter mass proxy (e.g. for
cosmological studies), but would give a different mass for the same
cluster. With PICACS we require both relations to give the same mass,
which results in larger scatter.

The covariance matrix $C_{T \mgas}$ was used to calculate the
Pearson's correlation between $\tint$ and $\mgint$, giving
$\rho_{T\mgas}=0.3\pm0.5$. Thus these data do not provide useful
constraints on the correlation of the scatter in these relations. The
constraints on the covariance matrix for the intrinsic scatter in the
VA sample are summarised in Table \ref{tab:cov}.

It should be noted that the clusters in the VA sample were selected to
be highly relaxed systems to permit reliable hydrostatic masses, so
the scatter values measured here and their covariance may not
represent the cluster population at large.

\begin{table}
\begin{center}
\begin{tabular}{l|cc}
  \hline
   & \tint & \mgint \\
  \hline
  \tint & $(1.5\pm1.2)\times10^{-3}$ & $(0.8\pm1.5)\times10^{-3}$  \\
  \mgint & $(0.8\pm1.5)\times10^{-3}$  & $(4.5\pm3.5)\times10^{-3}$  \\
  \hline
\end{tabular}
\caption{\label{tab:cov} Covariance matrix $C_{T \mgas}$ for the PICACS
fit to the VA data. The covariance was measured in $\log_{10}$ space.}
\end{center}
\end{table}

In addition to the covariance in the intrinsic scatter, we can also
investigate the degeneracies in the PICACS model parameters. The
correlations between parameters are shown in the scatterplot matrix in
Figure \ref{f.vamat}. This indicates that there is a mild degeneracy
between the two normalisation terms and the two slope terms in the
model, which is not surprising given the dependency of both scaling
relations on the cluster mass. Otherwise, no strong degeneracies exist
in the model in this case.

\subsection{Mass constraints}
\begin{figure}
\begin{center}
\scalebox{0.43}{\includegraphics*{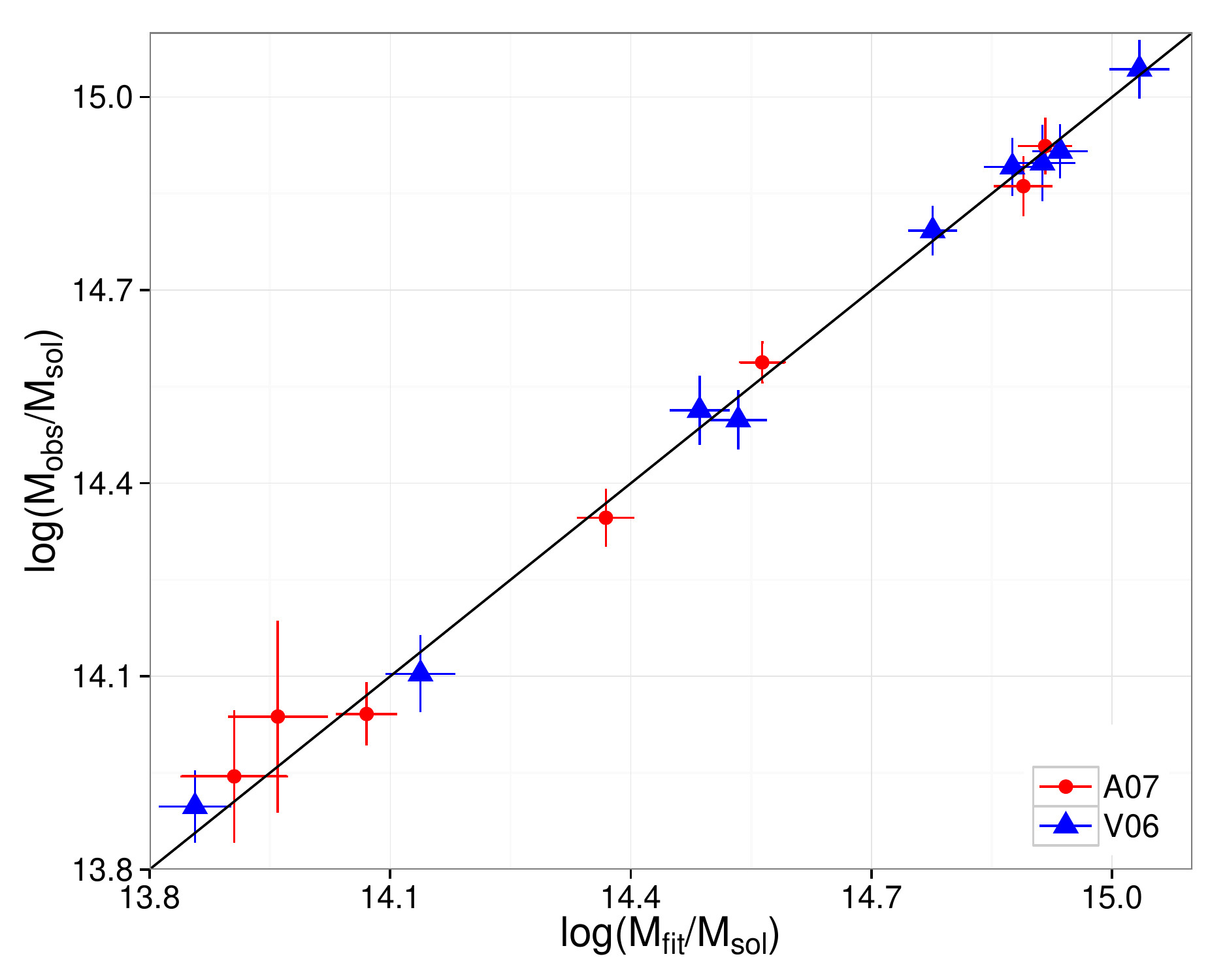}}
\caption[]{\label{f.vamm} X-ray hydrostatic \mobs\ is plotted against
  \mfit\ from the combined PICACS model for the VA sample. The line
  shows equality between the two mass estimates.}
\end{center}
\end{figure}

The best-fitting masses from the PICACS analysis are given in Table
\ref{tab:vamfit}, and are compared with \mobs\ in Figure
\ref{f.vamm}. The agreement is excellent, which should not be
surprising, given that \mfit\ is well-constrained by \mobs. It is
interesting to note that the uncertainties on \mfit\ are smaller than
those on \mobs; the median uncertainty on \mobs\ is $10\%$, while it
is $8\%$ on \mfit. This modest increase in precision comes from the
additional constraining power given by requiring the best-fit masses
to comply with the best-fit scaling relations in addition to being
constrained by the error bars on the observed hydrostatic masses.

\section{Application to Clusters Without Observed Masses}\label{sec:mfit}
We now apply the PICACS technique to the REXCESS sample, which is a
representative set of low-redshift clusters with high-quality \XMM\
data \citep{boh07}. The global X-ray properties ($T$, $L$, and \mgas)
of the clusters were presented in \citet[][hereafter P09]{pra09a} along with a study of
the luminosity scaling relations of the sample. X-ray hydrostatic
masses are not currently available for the sample. The representative
nature of the REXCESS sample, along with the limited redshift range
and precisely measured X-ray properties make it a good choice for a
second case study of the PICACS methodology. In the following, we use
the $T$, $L$, and \mgas\ values of P09, measured out to
\rf, with the central $15\%$ of \rf\ excluded for $T$ and $L$. As
before, the evolution parameters $\gamtm$ and $\gamma_L$ are fixed
at their self-similar values. The properties of the clusters used in
this study are summarised in Table \ref{tab:mfit}.

\begin{table*}
\begin{center}
\begin{tabular}{lcccccccccc}
  \hline
  Name & $z$ & $T$ & \mgas & $L$ & \mfit \\
       &   & keV & $10^{13}\Msol$ & $10^{44}\ergps$ & $10^{14}\Msol$ \\
  \hline
RXCJ0003$+$0203 & 0.092  & $3.83\pm0.10$ & $1.99\pm0.04$ & $1.15\pm0.01$ & $2.17 \pm 0.31$ \\
RXCJ0006$-$3443 & 0.115  & $5.24\pm0.20$ & $4.48\pm0.11$ & $3.17\pm0.05$ & $3.83 \pm 0.54$ \\
RXCJ0020$-$2542 & 0.141  & $5.54\pm0.13$ & $4.06\pm0.06$ & $4.05\pm0.03$ & $4.01 \pm 0.60$ \\
RXCJ0049$-$2931 & 0.108  & $2.87\pm0.10$ & $1.66\pm0.03$ & $0.99\pm0.02$ & $1.62 \pm 0.23$ \\
RXCJ0145$-$5300 & 0.117  & $5.81\pm0.15$ & $4.85\pm0.06$ & $3.87\pm0.03$ & $4.56 \pm 0.66$ \\
RXCJ0211$-$4017 & 0.101  & $2.08\pm0.05$ & $0.98\pm0.01$ & $0.48\pm0.01$ & $0.98 \pm 0.17$ \\
RXCJ0225$-$2928 & 0.060  & $2.53\pm0.14$ & $0.73\pm0.02$ & $0.31\pm0.01$ & $0.99 \pm 0.20$ \\
RXCJ0345$-$4112 & 0.060  & $2.28\pm0.07$ & $0.82\pm0.02$ & $0.37\pm0.01$ & $0.95 \pm 0.15$ \\
RXCJ0547$-$3152 & 0.148  & $6.04\pm0.14$ & $5.94\pm0.04$ & $5.73\pm0.04$ & $5.12 \pm 0.57$ \\
RXCJ0605$-$3518 & 0.139  & $4.93\pm0.12$ & $4.63\pm0.05$ & $4.23\pm0.03$ & $4.01 \pm 0.46$ \\
RXCJ0616$-$4748 & 0.116  & $4.17\pm0.11$ & $2.86\pm0.04$ & $1.88\pm0.02$ & $2.85 \pm 0.37$ \\
RXCJ0645$-$5413 & 0.164  & $7.23\pm0.18$ &$10.08\pm0.11$& $11.33\pm0.08$ & $7.73 \pm 0.98$ \\
RXCJ0821$+$0112 & 0.082  & $2.81\pm0.10$ & $1.16\pm0.03$ & $0.54\pm0.01$ & $1.23 \pm 0.17$ \\
RXCJ0958$-$1103 & 0.167  & $5.95\pm0.41$ & $4.43\pm0.20$ & $5.21\pm0.14$ & $4.48 \pm 0.78$ \\
RXCJ1044$-$0704 & 0.134  & $3.58\pm0.05$ & $3.32\pm0.04$ & $2.99\pm0.02$ & $2.68 \pm 0.56$ \\
RXCJ1141$-$1216 & 0.119  & $3.58\pm0.06$ & $2.45\pm0.02$ & $1.69\pm0.01$ & $2.26 \pm 0.29$ \\
RXCJ1236$-$3354 & 0.080  & $2.77\pm0.06$ & $1.21\pm0.02$ & $0.61\pm0.01$ & $1.31 \pm 0.18$ \\
RXCJ1302$-$0230 & 0.085  & $3.48\pm0.08$ & $1.80\pm0.02$ & $0.83\pm0.01$ & $1.75 \pm 0.27$ \\
RXCJ1311$-$0120 & 0.183  & $8.67\pm0.12$ &$10.69\pm0.06$& $14.93\pm0.07$ & $9.00 \pm 1.16$ \\
RXCJ1516$+$0005 & 0.118  & $4.68\pm0.10$ & $3.61\pm0.04$ & $2.76\pm0.02$ & $3.20 \pm 0.37$ \\
RXCJ1516$-$0056 & 0.120  & $3.70\pm0.09$ & $2.99\pm0.04$ & $1.77\pm0.02$ & $2.50 \pm 0.39$ \\
RXCJ2014$-$2430 & 0.154  & $5.75\pm0.10$ & $7.19\pm0.07$ & $7.47\pm0.06$ & $5.59 \pm 0.70$ \\
RXCJ2023$-$2056 & 0.056  & $2.72\pm0.09$ & $1.03\pm0.02$ & $0.40\pm0.01$ & $1.16 \pm 0.18$ \\
RXCJ2048$-$1750 & 0.147  & $5.06\pm0.11$ & $5.50\pm0.05$ & $4.40\pm0.03$ & $4.21 \pm 0.64$ \\
RXCJ2129$-$5048 & 0.080  & $3.84\pm0.14$ & $2.23\pm0.04$ & $1.19\pm0.02$ & $2.29 \pm 0.34$ \\
RXCJ2149$-$3041 & 0.118  & $3.48\pm0.07$ & $2.48\pm0.03$ & $1.58\pm0.01$ & $2.28 \pm 0.28$ \\
RXCJ2157$-$0747 & 0.058  & $2.79\pm0.07$ & $1.12\pm0.02$ & $0.37\pm0.01$ & $1.16 \pm 0.22$ \\
RXCJ2217$-$3543 & 0.149  & $4.63\pm0.09$ & $4.37\pm0.04$ & $3.69\pm0.03$ & $3.66 \pm 0.45$ \\
RXCJ2218$-$3853 & 0.141  & $6.18\pm0.20$ & $5.67\pm0.07$ & $5.55\pm0.06$ & $5.15 \pm 0.63$ \\
RXCJ2234$-$3744 & 0.151  & $7.32\pm0.12$ & $9.87\pm0.10$ &$12.28\pm0.10$ & $7.66 \pm 0.90$ \\
RXCJ2319$-$7313 & 0.098  & $2.56\pm0.07$ & $1.74\pm0.03$ & $0.97\pm0.01$ & $1.52 \pm 0.27$ \\
   \hline
\end{tabular}
\caption{\label{tab:mfit}X-ray properties of the REXCESS clusters
  taken from \citet{pra09a} with the addition of the cluster masses
  determined from the PICACS fits with VA priors. Where the
  uncertainties on the measured REXCESS properties were asymmetric,
  the mean value is used. Properties are determined within \rf, with
  the central $0.15\rf$ excluded for $L$ and $T$.}
\end{center}
\end{table*}
In the absence of \mobs\ for the REXCESS sample, priors are needed on
a subset of the PICACS scaling relation parameters to break the
degeneracy between \mfit\ and the scaling relation shape
parameters. Initially, we will maximise the use of the information
from the VA sample, and use the constraints on $\Atm$, $\Btm$, $\Amm$
and $\Bmm$ from the PICACS analysis summarised in Table \ref{tab:mobs}
(encoded as Gaussian priors in linear space for the $B$ terms and
$\log_{10}$ space for the $A$ terms). We also use the posterior
probability distribution on the covariance of $T$ and $\mgas$ from the
VA sample (Table \ref{tab:cov}) as Gaussian priors on the first
$2\times2$ elements of the full $3\times3$ $C_{T \mgas L}$ covariance
matrix. We refer to this set of priors as the ``VA priors''. Later
(\textsection \ref{sec:covar-degen-1}) we will review the success of
the PICACS method with weaker priors. As discussed in \textsection
\ref{sec:cautions-caveats}, the VA priors are not optimal for the
REXCESS sample, as the VA sample selected relaxed clusters
(necessitated by our use of hydrostatic masses), while the REXCESS
clusters encompass the full range of dynamical states.

\begin{figure*}
\begin{center}
\scalebox{0.29}{\includegraphics*{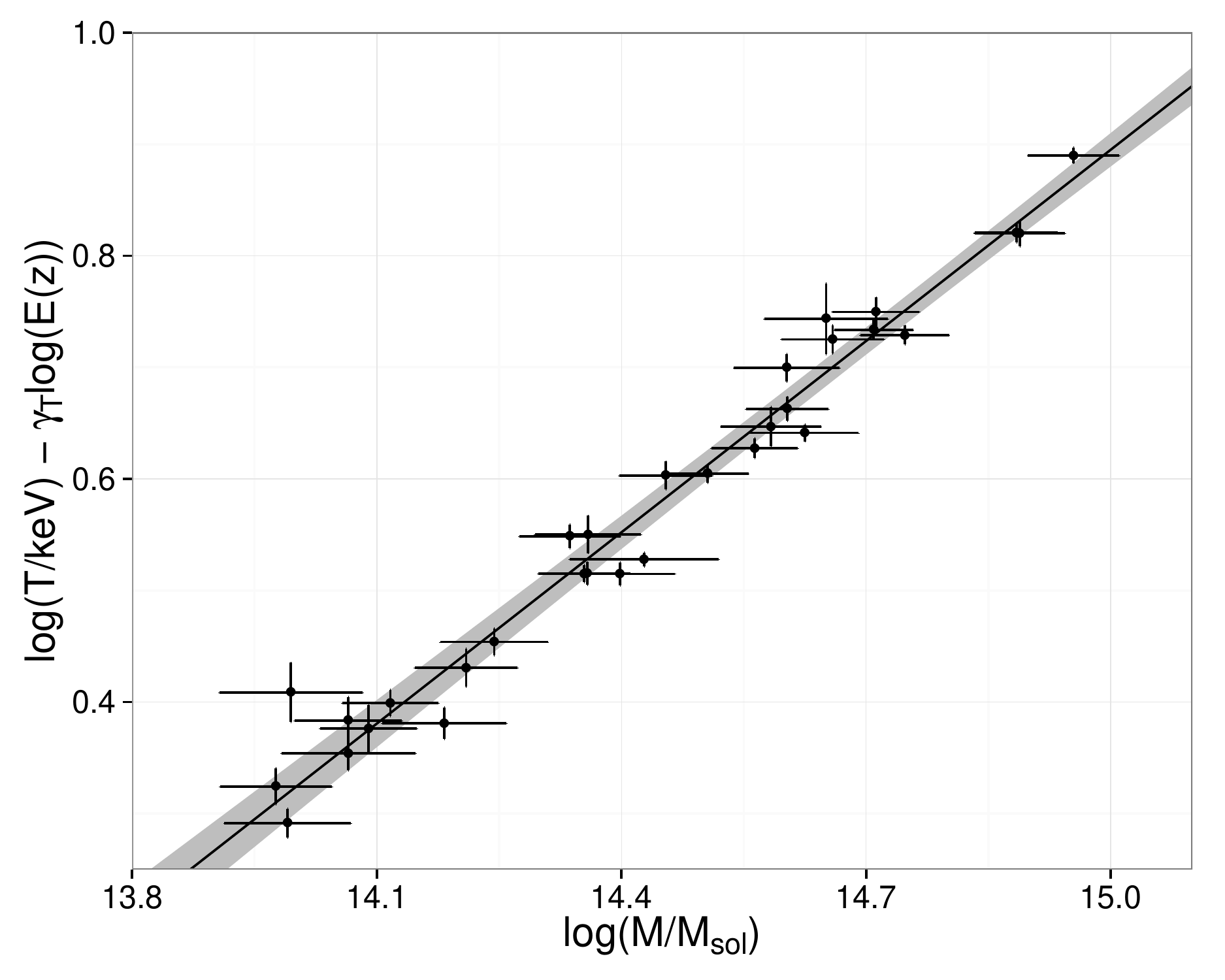}}
\hspace{0.1cm}
\scalebox{0.29}{\includegraphics*{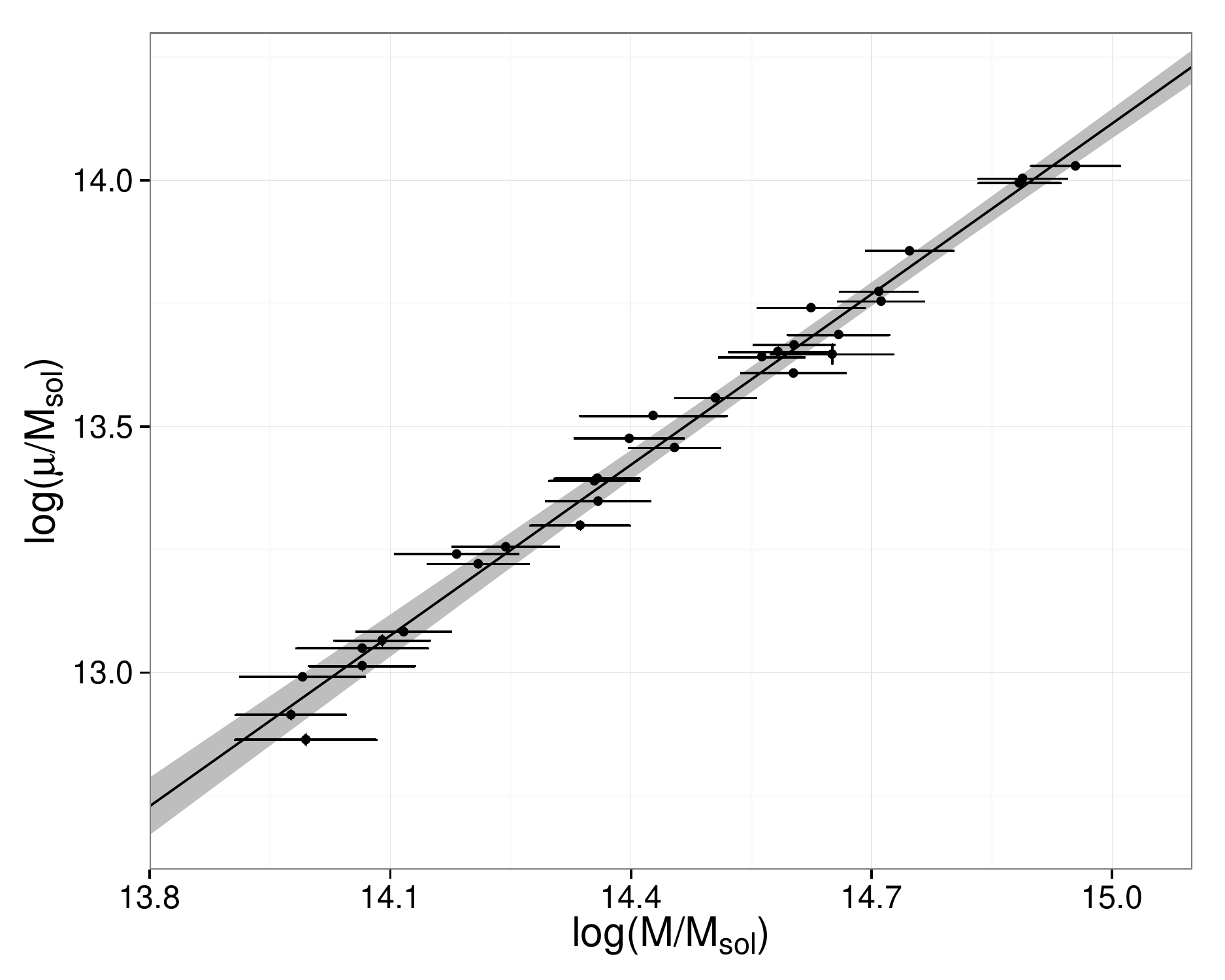}}
\hspace{0.1cm}
\scalebox{0.29}{\includegraphics*{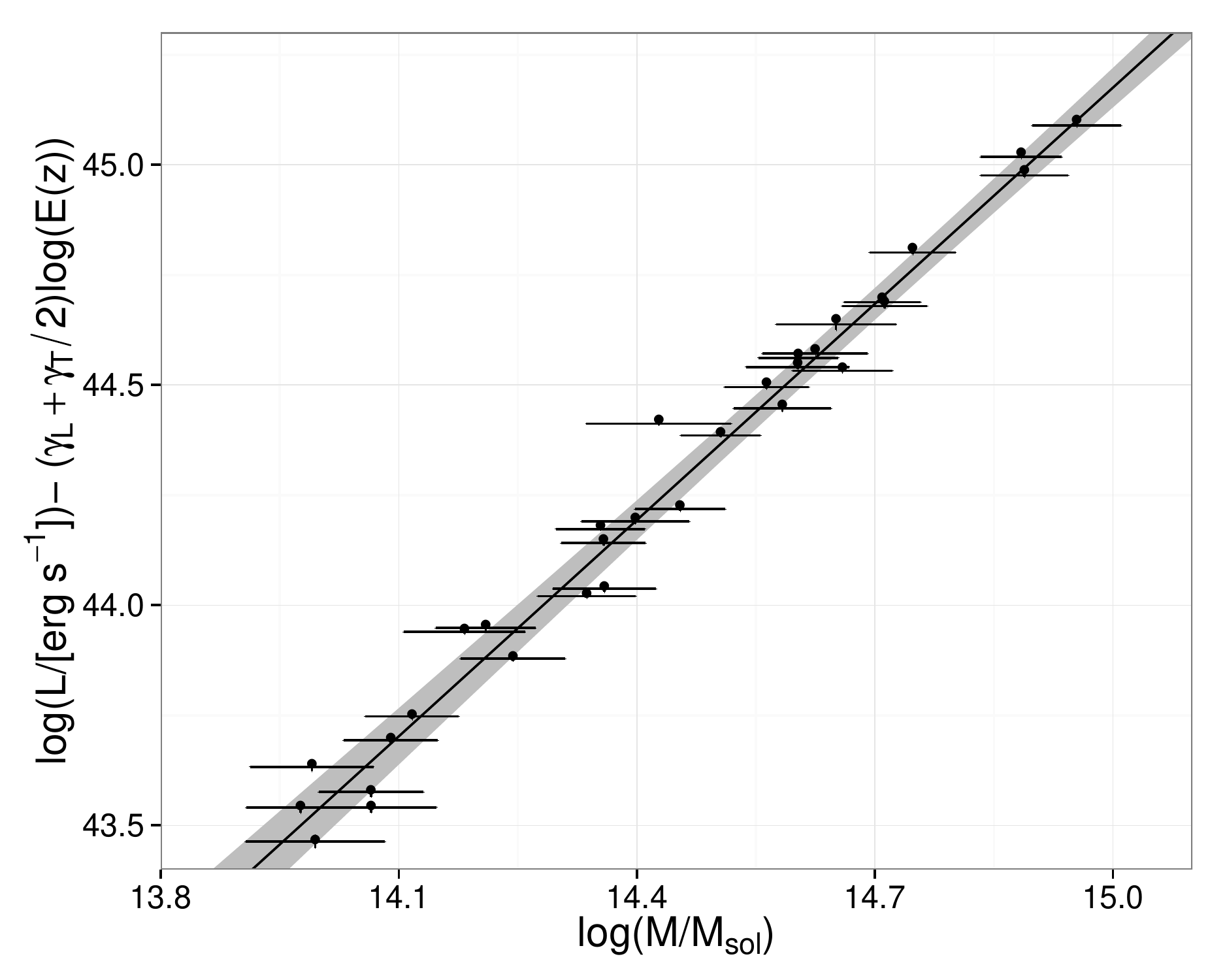}}
\caption[]{\label{f.rex} Best fitting $TM$ (left), $\mgas M$ (centre),
  and $LM$ (right) scaling relations for the REXCESS sample, with the
  shaded envelope indicating the $1\sigma$ uncertainty. The evolution
  parameters were fixed at their self-similar values,
  $\gamtm=2/3,\gamL=2$. Note that in these plots, the mass of each
  cluster is \mfit, the mass determined by the combined PICACS fit of
  the masses and scaling relations.}
\end{center}
\end{figure*}

The best-fitting PICACS relations to the REXCESS sample are plotted in
Figure \ref{f.rex}, and the parameters are summarised in Table
\ref{tab:rex}. These plots differ from those of conventional scaling
relations, as the masses plotted are the best-fitting masses from the
combination of scaling relations, and not observed masses.
The constraints on the scaling relation parameters are consistent with
the VA priors, but with slightly improved precision, and the new
constraints on the luminosity scaling parameters are quite precise.

The best-fitting PICACS relations can be compared with the REXCESS
$LM$ relation measured in P09. Rescaling to $M_0=2\times10^{14}\Msol$
and $L_0$ of unity as in P09, we find $A_{LM}=(1.08\pm0.11)\times
10^{44}\ergps$ and $B_{LM}=1.64\pm0.09$. While the normalisation
agrees well with P09, there is mild tension between the values of the
slopes ($A_{LM}=(1.08\pm0.04)\times 10^{44}\ergps$ and
$B_{LM}=1.80\pm0.05$), though we note that the comparison is not
exact, as P09 derive their $LM$ relation by using the $Y_XM$ relation
of A07 to convert their measured $Y_X$ values to masses.

Note that $\Bqm$ is consistent with the self-similar value of
1, which implies no mass scaling of the ICM structural parameter. This
is discussed in more detail later in the context of the $LT$
relation. We will now investigate the degeneracies in the model and
the sensitivity of the results to the choice of priors.

\begin{table*}
\begin{center}
\scalebox{0.9}{
\begin{tabular}{lccccccccccc}
\hline
Method & $\Atm$ & $\Btm$ & \inttm & $\Amm$ & $\Bmm$ & \intmm &
$\Alm$ & $\Blm$ & \intlm & $\Bqm$ \\
\hline
priors & $1.07\pm0.04$ & $0.59\pm0.04$ & $C_{T\mgas}$
       & $1.14\pm0.08$ & $1.24\pm0.07$ & $C_{T\mgas}$
       & - & - & - & -\\

fit    & $1.06\pm0.03$ & $0.57\pm0.03$ & $0.09\pm0.03$
       & $1.17\pm0.06$ & $1.16\pm0.06$ & $0.18\pm0.05$
       & $0.97\pm0.08$ & $1.64\pm0.09$ & $0.27\pm0.08$
       & $1.04\pm0.06$
       \\
\hline
\end{tabular}
}
\caption{\label{tab:rex} Best fitting parameters of the PICACS scaling
  relations fit to the REXCESS sample. All priors were
  encoded as Gaussian distributions with the specified mean and
  standard deviation, though the priors on the $A$ terms, while
  reported in linear space, were implemented as Gaussians in
  $\log_{10}$ space. For convenience, the intrinsic scatter terms are
  computed from the diagonal elements of of the covariance matrix in
  natural log space, so represent fractional scatter.}
\end{center}
\end{table*}

\begin{table*}
\begin{center}
\begin{tabular}{l | c c c}
  \hline
   & \tint & \mgint & \Lint \\
  \hline
  \tint & $(1.8\pm1.1)\times10^{-3}$ & $(1.1\pm1.1)\times10^{-3}$
  & $(2.1\pm2.0)\times10^{-3}$  \\
  \mgint & $(1.1\pm1.1)\times10^{-3}$ & $(6.6\pm3.6)\times10^{-3}$ & $(8.7\pm5.3)\times10^{-3}$ \\
  \Lint & $(2.1\pm2.0)\times10^{-3}$ & $(8.7\pm5.3)\times10^{-3}$ & $(1.5\pm0.9)\times10^{-2}$ \\
  \hline
    & $\rho_{T\mgas}$ & $\rho_{TL}$ & $\rho_{\mgas L}$ \\
  \hline
    & $0.31\pm0.30$ & $0.37\pm0.31$ & $0.85\pm0.14$ \\
  \hline
\end{tabular}
\caption{\label{tab:covrex} Covariance matrix $C_{T \mgas L}$ (upper section), and
  corresponding correlation coefficients (lower section) for the PICACS
fit to the REXCESS data. The off-diagonal $Q$ terms were fixed at
zero. The covariance was measured in $\log_{10}$ space.}
\end{center}
\end{table*}

\subsection{Covariance and Degeneracies}\label{sec:covar-degen-1}
The intrinsic scatter covariance matrix \Ctml\ from our reference fit
to the REXCESS data is summarised in Table \ref{tab:covrex},
along with the corresponding correlation coefficients. The data
  show weak evidence for moderate positive correlation between the
  scatter in $T$ and \mgas, and between the scatter in $T$ and
  $L$. There is strong evidence for a strong positive correlation in
  the scatter in \mgas\ and $L$. This is not surprising given the
  strong dependency of $L$ on \mgas, as illustrated in equation
  \eqref{eq:ltm}. This is comparable to the correlation between
  $\fgas$ and $L$ of $0.76$ found in the simulations of
  \citet{sta10}. This strong correlation demonstrates that the scatter
  in the $LM$ relation has a significant contribution from the scatter
  in the $\mgas M$ relation. A measurement of the covariance between
  observables is important, as it provides a means to model the
  propagation of biases due to X-ray flux based selection to other
  observables. For example, our results suggest that without taking
  this covariance into account, cluster masses estimated from \mgas\
  (or indeed $Y_X$) in an X-ray flux-limited sample would be biased
  high, with implications for cosmological studies using such
  techniques \citep{nor08,sta10,ang12}.

\begin{figure*}
\begin{center}
\scalebox{1.0}{\includegraphics*{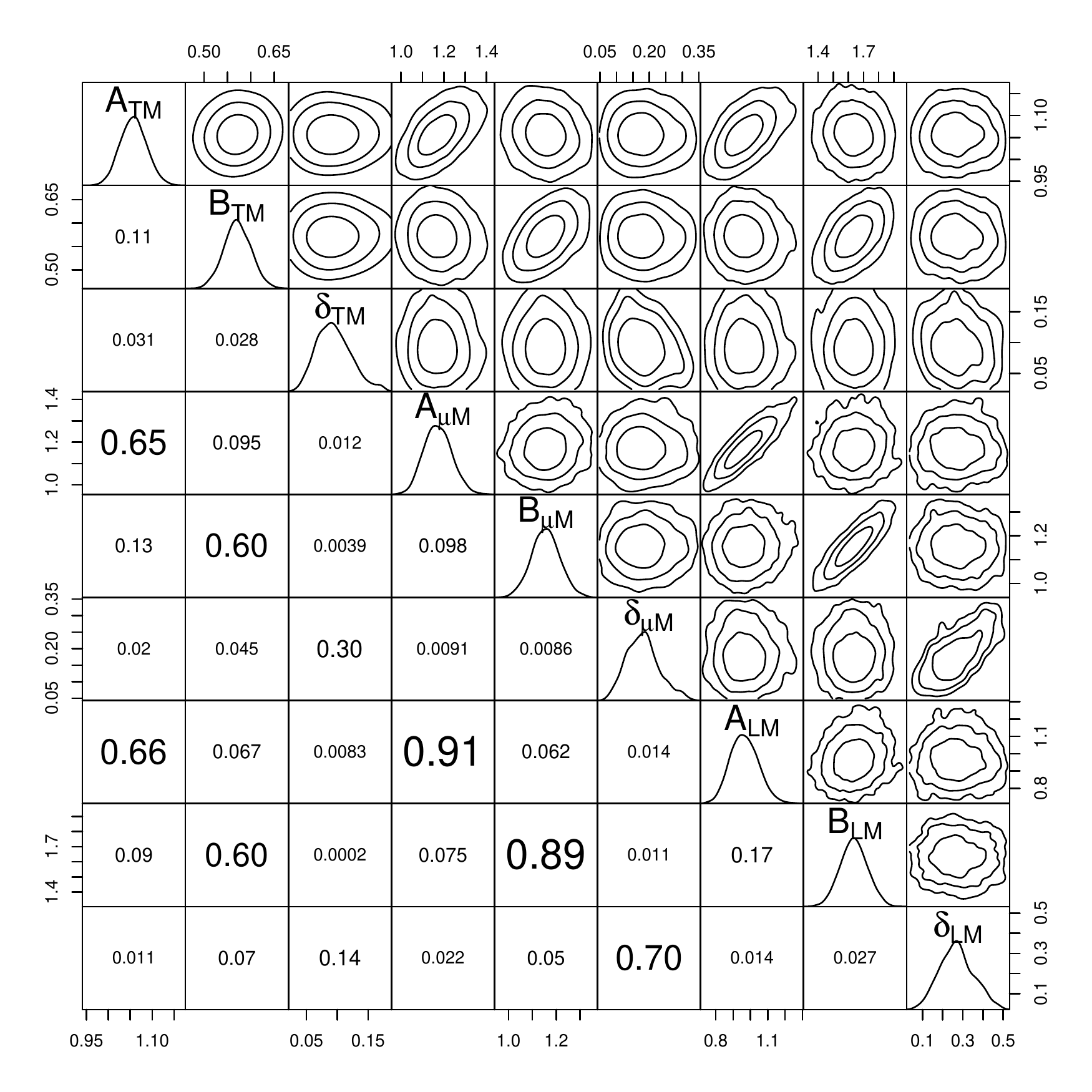}}
\caption[]{\label{f.rexmat} Correlation matrix of the PICACS model
  parameters for the fit to the REXCESS sample VA priors. The
  posterior densities are shown along the diagonal, with
    $1\sigma$, $2\sigma$, and $3\sigma$ confidence contours for the
    pairs of parameters shown on the upper triangle panels. The lower
  triangle panels show the magnitude of the Pearson's correlation
  coefficient for the corresponding pair of parameters (with a text
  size proportional to the correlation strength). The $\delta$ terms
  are in natural log space, and were computed from the square root of
  the diagonal elements of the covariance matrix \Ctml\ and so do not
  represent the full information in the covariance matrix.}
\end{center}
\end{figure*}

In Figure \ref{f.rexmat} the correlations of the model parameters for
the reference fit to the REXCESS data are plotted. Unsurprisingly,
strong degeneracies exist between the normalisations and between the
slopes of the scaling relations. This is due to the mass of each
cluster being a free parameter in each scaling relation. We also see
that
a degeneracy is present between the magnitudes of the scatter, \intmm\
and \intlm. As above, this is due to the strong dependency of the
observed luminosity on the baryon fraction, which is only partially
broken by the VA prior on the $T$ and \mgas\ terms of the $C_{T\mgas
  L}$ covariance matrix. Without additional information (e.g. from
\mobs) it is not possible to constrain \intmm\ and \intlm\
independently.

We investigated the sensitivity of the PICACS approach to the number
of priors by removing the VA priors on the $TM$ relation, but keeping
those on the $\mgas M$ relation, and on the $T,\mgas$ components of
\Ctml. In this case, the degeneracies between the slope parameters
seen in Figure \ref{f.rexmat} were stronger, and the fit did not
become formally stationary, with the slope parameters moving
coherently around on these lines of degeneracy. However, various
samples taken from the chain showed that all parameters remained
within $\approx1\sigma$ of their values when the full VA priors were
used. We thus recommend that priors on two of the three relations are
used for analyses where \mobs\ are not available for at least some of
the clusters.

\section{PICACS Mass Estimates}\label{sec:picacs-mass-estim}
A useful application of PICACS is the estimation of masses for
clusters without \mobs. The best fitting masses are automatically
estimated as part of the Bayesian inference process, and provide
masses that are fully consistent with the observed properties and
derived scaling relations. In \textsection \ref{sec:mobs}, we saw that
PICACS made a modest improvement to the precision of the hydrostatic
mass estimates for the VA sample. In this section we evaluate the
performance of PICACS at constraining the unknown masses of the
REXCESS sample. The best-fitting PICACS masses from our fits with VA
priors are given in Table \ref{tab:mfit}, and the median precision is
$14\%$.

The conventional way to estimate X-ray masses for a sample of clusters
such as this, in the absence of hydrostatic masses, is to use a single
scaling relation to estimate the mass from a single observable
(e.g. $T$, \mgas), simple combinations of observables such as $Y_X$ or
more generalised combinations of observables \citep{ett12}. Typically,
when doing this, only the statistical errors on the observable are
propagated to the mass estimate (\mfit), or at best, the uncertainties
on the shape parameters of the scaling relation are also
propagated. Generally, the contribution from the intrinsic scatter in
the relation is ignored, but this may be a significant contributor
when the statistical errors on the observable are small (e.g. for the
REXCESS sample, the median statistical error on \mgas\ is $1\%$). It
is straightforward, using a Bayesian approach to include the intrinsic
scatter and all of the uncertainties on the final mass estimate.

Let us define a generic scaling relation between mass and some
observable (or combination of observables) $X$. Using our previous
notation, we have
\begin{align}\label{eq:XM}
\frac{X}{X_0} & = A_{X} E(z)^{\gamma_{X}}\left(\frac{M}{M_0}\right)^{B_{X}}
\end{align}
and our likelihood function is
\begin{align}
\lik_X & = P(\xobs|\xint,\sigx)P(\xint|\xmod,\intx)\\
       & = P(\xobs|\xint,\sigx)P(\xint|M,\theta_X,\intx)
\end{align}
The posterior probability distribution of the model parameters is then
\begin{align}\label{eq:px}
P(\theta_X,M,\xint|\xobs) & \propto \lik_XP(\theta_X)P(M)P(\xint)
\end{align}
The priors on $\theta_X$ (denoting $A_X, B_X, \gamma_X$) and \intx\
are taken from the scaling relation to be applied. In most cases,
$\gamma_X$ will be fixed (i.e. at a self-similar value) and \intx\ may
not have measurement errors, though neither of these factors are
limitations of the Bayesian approach. In equation \eqref{eq:px}, we do
not expect significant additional constraints to be placed on
$P(\theta_X|\xobs)$ or $P(\xint|\xobs)$, but $P(M)$ will be jointly
constrained by the priors on those terms and by \xobs, and fully
marginalised over all of the uncertainties.

We apply this method to compare the precision of the PICACS mass
estimates for the REXCESS sample with those obtained from a single
scaling relation. We use the $Y_X M$ relation of A07, for which the
intrinsic scatter was given as $0.039$ in log$_{10}$ space, with no
uncertainties provided. We convert this to the intrinsic scatter in
$Y_X$ by dividing by the slope of the A07 $Y_X M$ relation, and
transform to natural log space to give $\delta_{Y_X}=0.16$. Including
this intrinsic scatter, and the errors on $A_{Y_X}$, $B_{Y_X}$ and
$Y_X$, we find a median uncertainty on the REXCESS masses of
$10\%$. Neglecting the intrinsic scatter results in a median mass
precision of $4\%$, while including an uncertainty on $\delta_{Y_X}$
of the form $\delta_{Y_X}=0.16\pm0.05$ (a reasonable estimate based on
our fits to the VA data) slightly increases the median uncertainty to
$11\%$. Recall that the median PICACS mass uncertainty for the same
clusters was $14\%$. We thus find that in the absence of \mobs, PICACS
provides mass estimates of slightly poorer precision
compared to a single scaling relation, but has the advantage of
providing masses that are simultaneously consistent with all of the
observables.

As discussed in \textsection \ref{sec:covar-degen}, the intrinsic
scatter measured with PICACS is typically larger than that measured
for single scaling relations with traditional fitting techniques. Thus
single or composite scaling relations that are optimised to reduce the
intrinsic scatter \citep[such as $Y_X$, or the generalised scaling
relations of][]{ett12,ett13a} may provide higher precision mass
estimates. However, such techniques are less useful than PICACS for
studying the astrophysics that shape the scaling relations.

\section{The PICACS $Y_XM$ Relation}\label{sec:picacs-tm-relation}

\begin{figure}
\begin{center}
\scalebox{0.43}{\includegraphics*{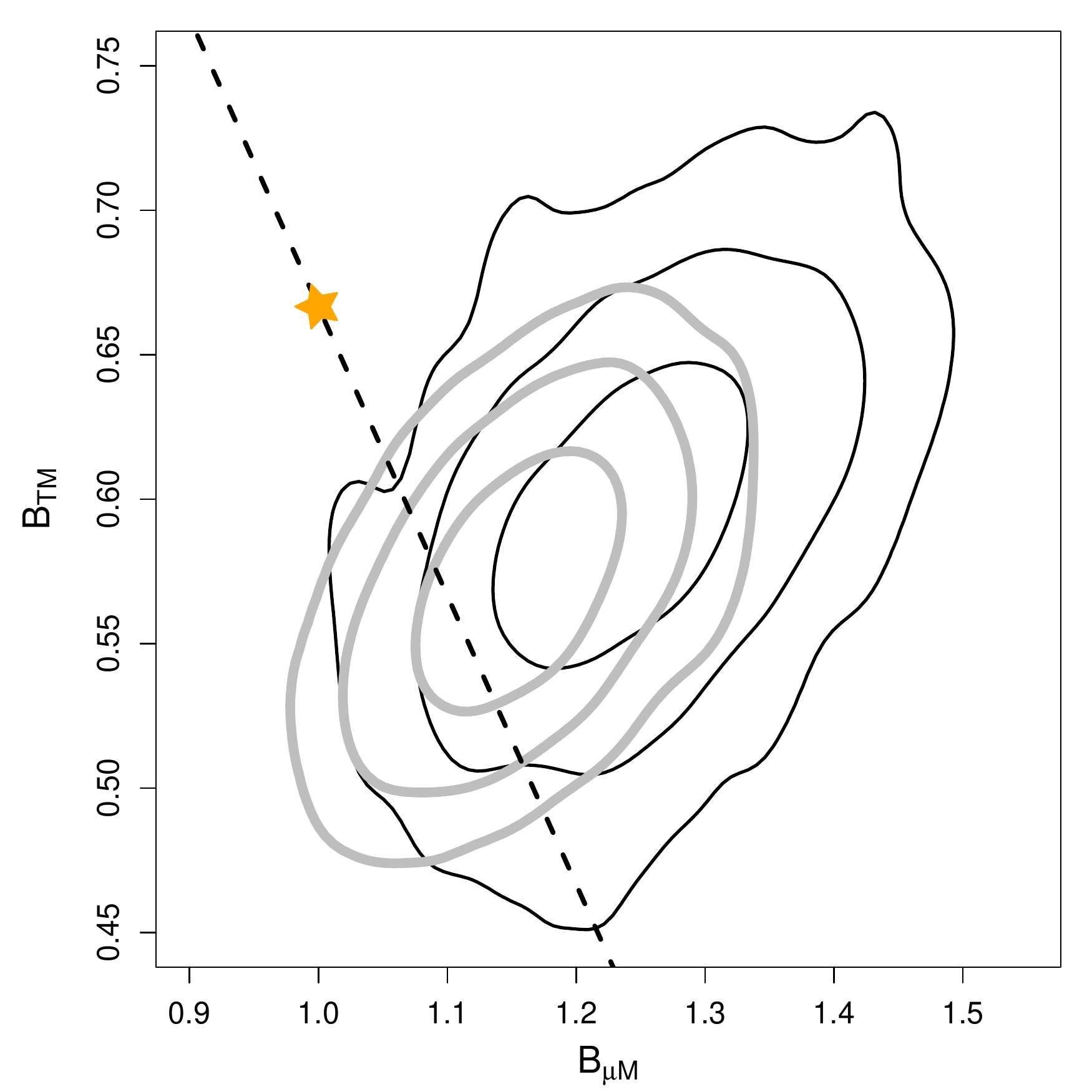}}
\caption[]{\label{f.bmbtm} Posterior probability contours for $\Bmm$
  and $\Btm$ for the fit to the VA data (narrow black contours) and
  the fit to the REXCESS data with VA priors (thick grey
  contours). Contours are set at the $1\sigma$, $2\sigma$ and
  $3\sigma$ levels. The dashed line shows the locus of the
  self-similar $Y_XM$ relation ($\Bmm+\Btm=5/3$), and the star
  marks the self-similar values of $\Bmm=1$, $\Btm=2/3$.}
\end{center}
\end{figure}

The constraints provided by the PICACS model on the $Y_XM$ relation
are illustrated in Figure \ref{f.bmbtm} shows the posterior
probability contours from the VA and REXCESS fits in the $\Bmm-B_{TM}$
plane, along with the line corresponding to the self-similar $Y_XM$
relation ($B_{TM}+\Bmm=5/3$). Both fits to the fundamental $TM$ and
$\mgas M$ relations are inconsistent with self-similarity, but are
close to the locus of the self-similar $Y_XM$ relation (though recall
that the fits are not independent - the VA fit provided the priors for
the REXCESS fit).

This is consistent with the suggestion of A07 that the thermal energy
content of the ICM, as represented by $Y_X$, is the quantity most
closely related to the cluster mass. The low scatter observed in the
$Y_XM$ relation \citep[][A07]{kra06a,mau07b} implies that the ICM in
clusters of a given mass has very similar total thermal
energy. Meanwhile, the coordination of the slopes of the $\mgas M$ and
$TM$ relations to maintain a close to self-similar slope of the $Y_XM$
relation while being far from self-similar themselves, implies that
the mechanism responsible for depletion (or preventing accretion) of
the ICM in lower mass clusters also results in an increased
temperature of the remaining gas.

This is compatible with models in which feedback preferentially
removes low-entropy gas from the ICM (by removal, we mean that the gas
is moved out beyond, or prevented from accreting within, \rf), and
does so more effectively in low mass systems
\citep[e.g.][]{voi05b,mcn07,pra10,mcc11}. The generic result of this
feedback is increasing depletion of the ICM within \rf\ in lower mas
halos, with the remaining higher entropy gas having a temperature
consistent with the virial temperature, due to its longer cooling
time. However, this does not complete the picture, as a steeper than
self-similar $TM$ relation combined with a self-similar $Y_XM$
relation would require that the remaining gas is heated by an amount
equivalent to the thermal energy lost by the low entropy gas as it
cooled and was removed. Furthermore, this heating must affect the mean
temperature of the gas outside the central regions ($0.15\rf$), which
were excluded in the temperatures used for this study.

These results should be treated with some caution, though, as the VA
data alone are agnostic as to whether it is the $TM$ or $Y_XM$
relation that is self-similar. It is only when the REXCESS data are
analysed with VA priors that the self-similar $Y_XM$ relation is
strongly preferred, but as discussed in \textsection
\ref{sec:cautions-caveats}, the VA priors are not optimal for the
REXCESS sample. In particular, the representative nature of the
REXCESS sample means that it will encompass a broader range of
feedback states than the relaxed clusters in the VA sample. The most
robust results will come from the analysis of representative samples
with direct observational constraints on the total masses.

There is some variation in the literature in recent studies of the
slopes of these relations. For example, A07 find a shallower $TM$ and
steeper $Y_XM$ relative to the self-similar values (their best fitting
values are very close to our fit to the combined VA data), while
\citet{vik09} find slopes of both relations that are consistent with
being self-similar. However, the results agree at the $1\sigma$ level,
and we argue that a combined analysis of the $TM$ and $\mgas M$
relations as presented here is the most useful way to investigate the
physical processes shaping these relations.

\section{The PICACS $LT$ Relation}\label{sec:picacs-lt-relation}
\begin{figure}
\begin{center}
\scalebox{0.43}{\includegraphics*{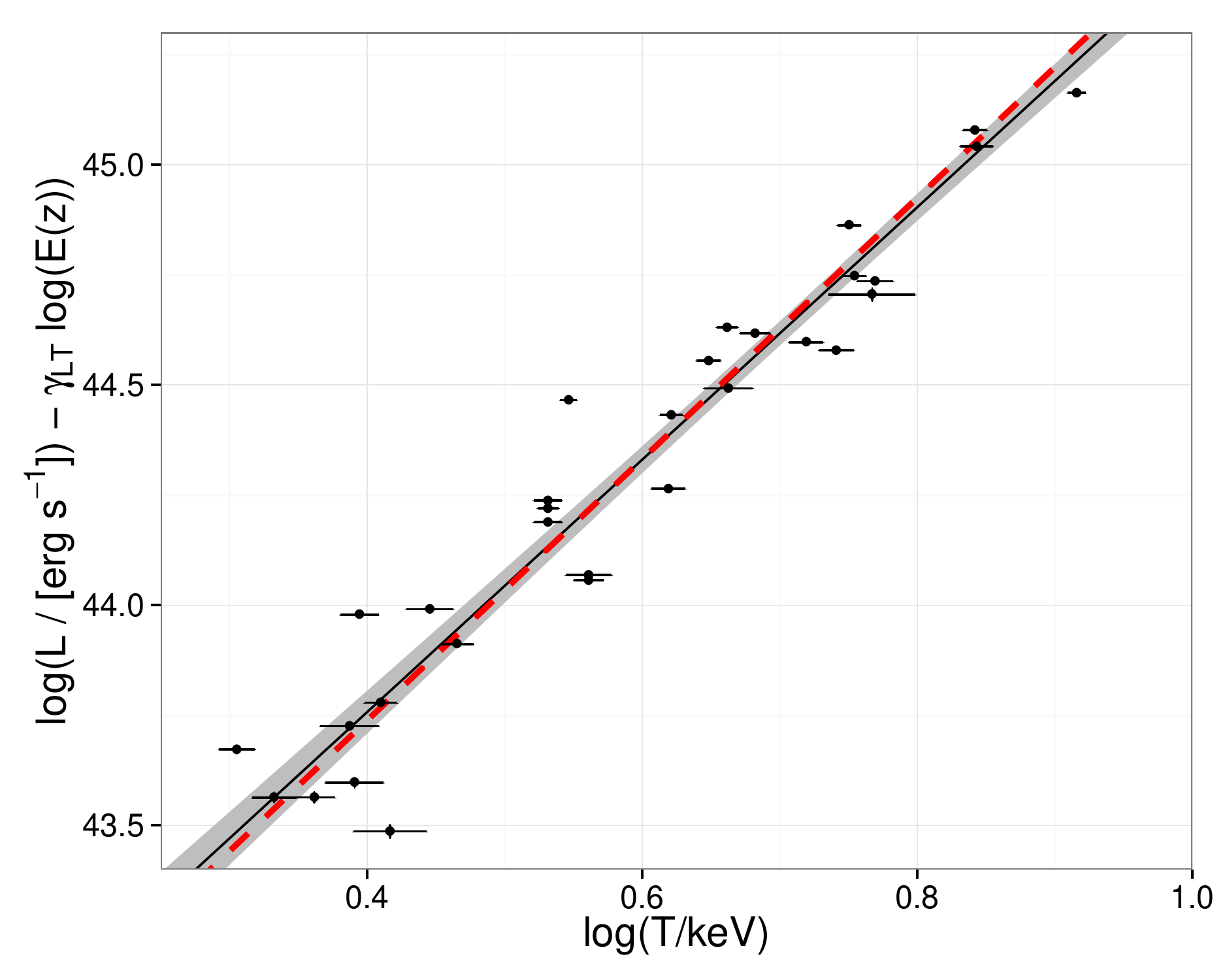}}
\caption[]{\label{f.lt} The $LT$ relation predicted by the PICACS fit
  to the REXCESS mass scaling relations is plotted with the observed
  REXCESS properties. The solid line shows the PICACS model, which is
  not fit directly to the data in this $LT$ plane, with the grey
  envelope giving the $1\sigma$ uncertainty. The dashed line shows the
  BCES orthogonal regression as fit to the data in this
  plot. Luminosities are scaled by the best fitting PICACS
  $\gamma_{LT}$ evolution parameter (see text for details), and $L$
  and $T$ were measured in the $[0.15-1]\rf$ aperture.}
\end{center}
\end{figure}

We will now examine the $LT$ relation predicted by the PICACS fit to
the REXCESS data. Recall that we do not fit the data directly in the
$LT$ plane, but the form of the $LT$ relation is given by the
 self-consistent PICACS models (see equations \eqref{eq:alt},
\eqref{eq:blt}, \eqref{eq:glt}). In Figure \ref{f.lt} we plot the
REXCESS data in the $LT$ plane, along with the PICACS $LT$
relation. Also plotted is the BCES orthogonal regression fit to the
data in the $LT$ plane. The REXCESS luminosities were scaled by PICACS
evolution parameter $\gamma_{LT}$ for the plot, and this is included
in the BCES fit (it is automatically part of the PICACS fit). Note
that in the current study, $\gamma_{LT}$ is not fit to any
redshift-dependence of the observed cluster properties, but gives the
expected self-similar evolution of the $LT$ relation when the
dependency on the slopes of the mass observable scaling relations in
equation \eqref{eq:glt} is included. The agreement between the PICACS $LT$
relation and the BCES fit is excellent; the parameters of the $LT$
models are summarised in Table \ref{tab:lt}.

\begin{table}
\begin{center}
\begin{tabular}{lccc}
\hline
Method & $A_{LT}$ & $B_{LT}$ & $\gamma_{LT}$ \\
\hline
PICACS & $0.82\pm0.05$ & $2.87\pm0.13$ & $0.42\pm0.09$ \\
BCES & $0.84\pm0.05$ & $2.97\pm0.16$ & $0.42^\dagger$ \\
\hline
\end{tabular}
\caption{\label{tab:lt} Best fitting parameters of the $LT$ relation
  of the REXCESS sample predicted by the PICACS fit and determined
  from a BCES orthogonal regression to the $LT$ data. The evolution
  parameter $\gamma_{LT}$ does not include any redshift-dependence of
  the cluster properties, but gives the self-similar evolution
  including the dependency on the slopes of the mass observable
  scaling relations in equation \eqref{eq:glt}. $^\dagger$The
  luminosities were scaled by this fixed value for
  the BCES fit. See the text for a full discussion of the evolution
  parameter.}
\end{center}
\end{table}

As implied by the agreement with the BCES fit, the PICACS LT relation
is also consistent with the P09 fit to the same REXCESS data. The P09
slope of $2.94\pm0.15$ agrees very well with the PICACS slope in table
\ref{tab:lt}. Rescaling to $L_0$ of unity, as in P09, the PICACS
normalisation at $5\keV$ is $A_{LT}=(4.11\pm0.26)\times10^{44}\ergps$,
compared with $A_{LT}=(4.06\pm0.22)\times10^{44}\ergps$ in P09, also
in excellent agreement. Note that the comparison is not exact, as the
PICACS fit incorporates correctly the self-similar evolution implied
by the slopes of the scaling relations, while in P09 the luminosities
are scaled by the traditional self-similar $E(z)^{-1}$. In practice,
for this low redshift sample, the differing evolution corrections are
negligible.

In fact, while it is reassuring that the PICACS
method is able to reproduce the observed $LT$ relation, this should
not surprise us; it simply demonstrates that the three PICACS scaling
relations form an internally consistent description of the observed
properties. Note that the good agreement with the observed $LT$
relation does not necessarily indicate that the individual scaling
relations are a good description of the clusters. For instance, if no
prior is included on the slope of any of the scaling relations, the
degeneracy of the slopes leads to unphysical values for \mfit\ and the
slope parameters. However the internally-consistent scaling relations
means that the combination of $A$ and $B$ parameters remains such that
the observed $LT$ relations is still reproduced reasonably well. In
other words, in the PICACS framework, the observed form of the $LT$
relation is a necessary consequence of requiring the observables to be
related to the same masses through power law relations, but is not
sensitive to the form of those relations.

\subsection{The slope of the $LT$ relation}
The advantage of the PICACS method is that while the BCES fit simply
tells us that the slope of the $LT$ relation is steeper than the
self-similar expectation of $B_{LT}=2$, PICACS enables us to decompose
this into the separate mass scaling relations. Table \ref{tab:rex}
shows that $\Bqm$ is consistent with unity. Recall that this parameter
describes the additional steepening of the luminosity mass relation,
beyond that due to the mass dependency of $T$ and \mgas, which we
ascribe predominantly to trends in the ICM structure with mass. The
PICACS results thus show that (given the VA priors) the steep slope of
the REXCESS $LT$ relation is consistent with the departures from self
similarity in $T$ and \mgas\ alone, with no additional contribution
from $Q$. This indicates that the ICM structure parameter $Q$ has no
mass dependency. While previous work has shown a significant mass
dependence of the shape of ICM surface brightness or density profiles
\citep[e.g.][]{san03,cro08,mau12}, the dominant effect of those
structural trends, when considering a relatively large region such as
\rf, is reflected by the mass-dependence of \fgas. Our results imply
that any additional contribution from a mass dependence of $Q$ is not
significant. A similar conclusion on the lack of mass dependency of
$Q$ was reached by P09, who estimated $Q$ directly from the REXCESS
gas density profiles.

\begin{figure}
\begin{center}
\scalebox{0.43}{\includegraphics*{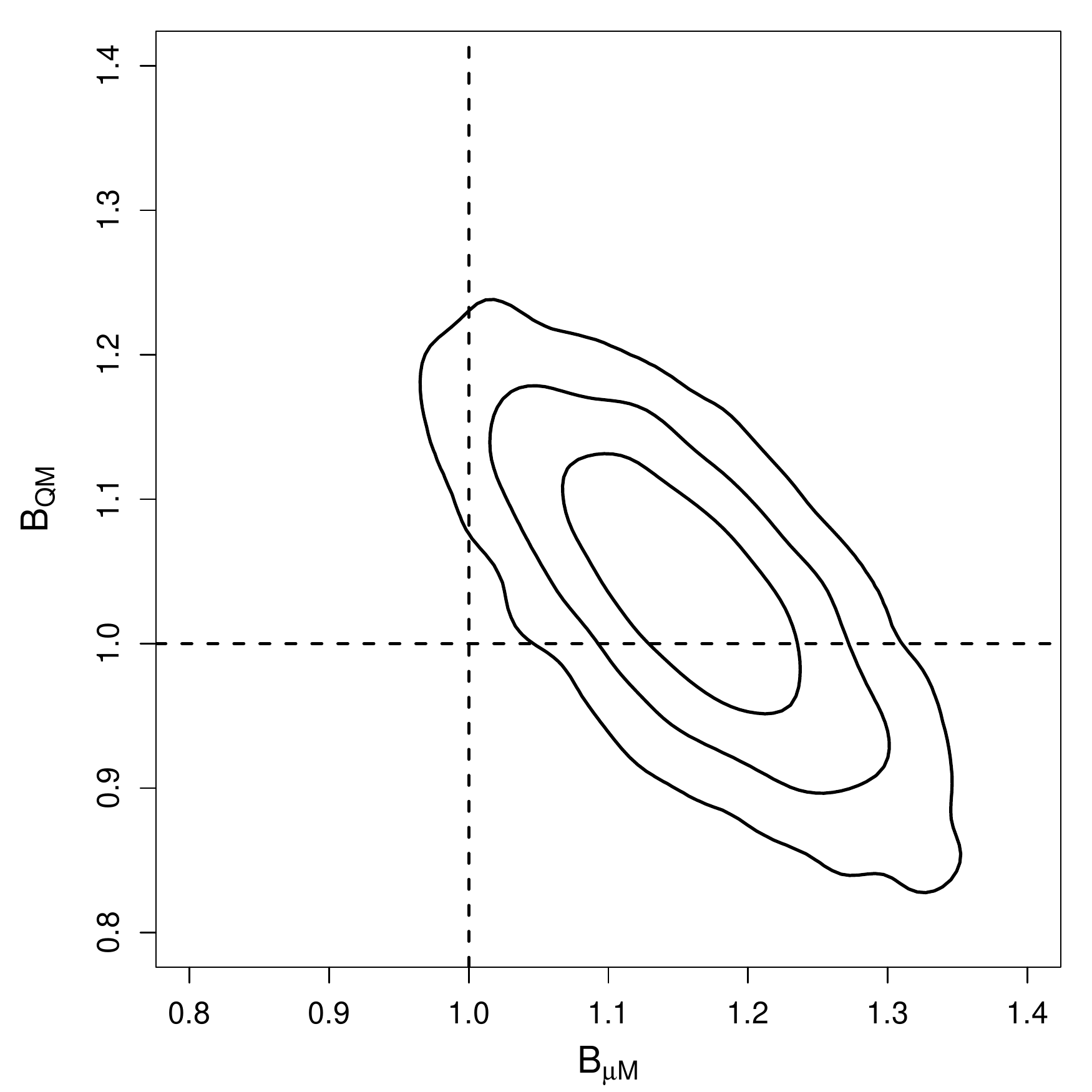}}
\caption[]{\label{f.bmbl} Posterior probability contours for $\Bmm$
  and $\Bqm$ for the REXCESS fit with VA priors. Contours are set at
  the $1\sigma$, $2\sigma$ and $3\sigma$ levels and the dotted lines
  shows the self-similar values of $\Bmm=1$, $\Bqm=1$.}
\end{center}
\end{figure}

This suggests that the $LT$ (and $LM$) relation is predominantly
shaped by the same process of gas removal and heating that shaped the
$TM$ and $\mgas M$ relations. The changes to the $TM$ and $\mgas M$
relations act in the sense of steepening the $LT$ relation such that
there is no need for additional influence from the $Q$ parameter.
This conclusion should, however, be treated with a little caution due
to the model degeneracies. Figure \ref{f.bmbl} shows the confidence
contours from the PICACS fit in the $\Bmm,\Bqm$ plane; taking the
parameter degeneracy into account, the data can only exclude $\Bmm=1$
at the $\sim2\sigma$ level. Furthermore, as we have seen, the breaking
of the degeneracy between $\Bmm$ and $\Bqm$ is sensitive to the choice
of prior.

\subsection{The Evolution of the $LT$ relation}
The standard approach to testing evolution of the $LT$ relation is to
use equation \eqref{eq:lt} and define $\gamma_{LT}=1$ as the reference
point for self-similar evolution \citep[e.g.][]{mau06a}. However, this
only holds true if the slopes of all of the mass scaling relations of
$T$, \mgas\ and $L$ are self-similar. In the event that they are not,
which has been suggested by many observational studies, then the
expected self-similar evolution of the $LT$ relation is not
$\gamma_{LT}=1$ but is given by equation \eqref{eq:glt}. Of course,
this does not imply that the slopes of the scaling relations influence
the evolution of clusters, it is simply a consequence of algebraic
manipulations used to derive the $LT$ relation.

If applied to a sample covering a significant redshift baseline,
PICACS can be used to fit the evolution of all scaling relations
self-consistently with their slopes, providing a true measurement of
their evolution. We reserve this investigation for a future paper, as
measuring the evolution of the scaling relations also requires
modelling of sample selection functions to avoid biases masking or
mimicking real evolution. This is quite possible within the PICACS
framework, along the lines laid out by \citet{man10a}.

For the current study, we simply note that the PICACS fit to the
REXCESS data gives the self-similar evolution of the $LT$ relation as
$\gamma_{LT}=0.42\pm0.09$, significantly weaker than the naive
expectation of $\gamma_{LT}=1$. Note again that the evolution
  parameters of the scaling relations were fixed at their self-similar
  values ($\gamtm=2/3,\gamL=2$ and no \fgas\ evolution); our
  measurement of $\gamma_{LT}=0.42\pm0.09$ is not a measurement of the
  evolution in the REXCESS data, it is a revised prediction of the
  self-similar evolution due to the non-self-similar slopes of the
  REXCESS scaling relations. This should be taken into account when
establishing a reference self-similar evolution against which to
measure deviations. For example, using this PICACS reference for the
self-similar $LT$ evolution reduces the significance of the weaker
than self-similar (or negative) evolution measured by recent studies
\citep{rei11,hil12}. Those results remain statistically significant
compared to our weaker self-similar reference, but the most robust
measurements of the evolution will come from a full PICACS analysis of
the cluster population to high redshift.

\section{Cautions and Caveats}\label{sec:cautions-caveats}
As is clear from Figure \ref{f.rexmat}, strong degeneracies exist in the
PICACS model when there are not direct observational constraints on
the cluster masses. Not shown in the correlation matrix are the
degeneracies between the scaling relation parameters and the fitted
cluster masses. These are mitigated with the use of priors on the
scaling relations, but without any informative priors, the degeneracy
is total; it would be quite possible for masses to fit to unphysical
values, and for the normalisations and slopes and scatters of the
relations to adjust to compensate. Without \mobs\ for at least a
subset of clusters, the PICACS fits are highly dependent on the choice
of priors for a subset of the scaling relation shape parameters.

In the case of our analysis of the REXCESS sample, the VA priors were
derived from a sample of relaxed clusters, while the REXCESS clusters
encompass a representative range of dynamical states. This difference
is unavoidable if X-ray hydrostatic masses are to be used as \mobs,
but could give rise to systematic effects in the derived REXCESS
scaling relations and e.g. our conclusions on the relative
contributions of the mass scaling relations to the steepening of the
$LT$ relation. The most robust PICACS analysis of representative
samples like REXCESS would require mass constraints from a techniques
such as gravitational lensing or caustic analyses which are
insensitive to cluster dynamical state.

A related complication is the dependence of the observed hydrostatic
mass on $T$ and \rhog. This introduces covariance between \mobs\ with
respect to the true mass and the other X-ray observables. This is not
addressed in our model, and so will influence our estimate of the
covariance in the VA sample, where we are effectively assuming no
intrinsic scatter between the hydrostatic \mobs\ and the true
mass. Furthermore, our analysis also assumed that the log-normal
intrinsic scatter covariance matrix was constant as a function of
cluster mass. This may well not be the case, as it is clear that
non-gravitational processes have an increasing effect on the gas
properties of lower mass systems. These issues are also best addressed
by using mass estimates that are independent of the X-ray data.

The determination of the covariance in the intrinsic scatter in the
cluster population depends crucially on the size of the uncertainties
on $L$, $T$, \mgas\ and \mobs. The errors on these quantities are
generally quoted as statistical errors only, but in fact there may be
significant systematic uncertainties on those quantities too. For
example, there remain calibration uncertainties for both \Chandra\ and
\XMM\ affecting all measured X-ray properties, and choices made during
the reduction and analysis of the data (e.g. data cleaning, background
treatment) also contribute. Hydrostatic masses can be influenced by
the method used for modelling the density and temperature profiles,
and whether a parametric form is assumed for the mass profile (see
e.g. the appendix of V06). In the analysis of the VA sample, we
allowed for systematic calibration offsets between $T$ and \mgas\
measured with \Chandra\ and \XMM\ with a simple multiplicative factor,
which turned out to be negligible. However, if there are any other
contributions from systematics to the uncertainties on the observed
quantities, that are not included in the errors quoted in V06, A07 and
P09, then the our determination of the intrinsic covariance will be
overestimated.

In its current form, PICACS does not include several factors which
could affect the measured scaling relations and masses. The most
significant of these is the modelling of Malmquist and Eddington
biases \citep[see e.g.][for a discussion in the context of scaling
relations]{all11a}. These biases can affect both the shape and
evolution of the mass scaling relations in X-ray selected samples. The
principal effect is a bias towards clusters with higher than average
luminosity for a given mass. Full treatment of these effects require
knowledge of the survey selection function and the mass function
describing the population from which the clusters were
sampled. \citet{man10a,man10b} have demonstrated how to include this
in a self-consistent analysis of cosmological parameters and scaling
relations. Extending PICACS along those lines will enable us to remove
any effects of bias in luminosity in the current analysis, while the
modelling of covariance between the scatter terms provides a natural
way to propagate the effects of the bias through to the other scaling
relations. For the current study, the REXCESS selection function is
known, but the VA sample has no selection function (due to the
cherry-picking of relaxed clusters for hydrostatic masses). This means
that a bias correction could only be approximate as the PICACS
analysis of the REXCESS clusters depends strongly on the VA priors.

We also currently do not model the uncertainty on the measurement
errors \citep[as in][]{and10}. This is not expected to have a large
effect on the current results due to the relatively high precision on
the observables, but could be important when modelling data with
larger measurement errors, and could plausibly affect the
determination of the magnitude and covariance of the intrinsic scatter.

\section{Summary and Conclusions}\label{sec:summary-conclusions}
We have introduced PICACS, an internally consistent physical model for
the analysis of galaxy cluster mass scaling relations, and a Bayesian
framework with which to implement it. PICACS provides
a self-consistent set of constraints on the parameters describing the
shape, scatter and evolution of the scaling relations and on the
masses of individual clusters. It may be used to study the scaling
properties of clusters with observed masses, estimate the masses of
clusters without observed masses, or a combination of the two. The new
method was demonstrated on several observational datasets, and the key
results were as follows:

\begin{itemize}
\item A PICACS analysis of the VA sample of relaxed clusters with
  precise X-ray hydrostatic masses was used to measure the shape and
  scatter of the $TM$ and $\mgas M$ scaling relations, producing
  result in excellent agreement with traditional regression methods.
\item Our analysis of the REXCESS sample of clusters, which lacks
  hydrostatic mass estimates, utilised priors from the VA analysis and
  was able to jointly constrain the scaling relations of $T$, \mgas\
  and $L$ with mass, and provide mass estimates for the clusters.
\item For the REXCESS sample with VA priors, the slopes of the $\mgas
  M$ and $TM$ relations were found to be significantly steeper and
  shallower, respectively, than the self-similar predictions, while
  their combination remains close to the self-similar slope of the
  $Y_XM$ relation. We interpret this as due to AGN feedback removing
  low-entropy gas from lower mass clusters, while heating the
  remaining gas, keeping the total thermal energy content of the ICM
  roughly constant.
\item The PICACS analysis of the REXCESS sample showed that the steep
  observed slope of the $LT$ relation is due solely to those changes
  in the $\mgas M$ and $TM$ relations, with no significant
  contribution due to structural variations of the ICM inside \rf\
  (i.e. no mass dependence of $Q$).
\item The PICACS framework fully accounts for the effect of the
  scaling relation slopes on the expected self-similar evolution of
  the $LT$ relation, and we show that the expected evolution is
  significantly weaker than is usually assumed when this effect is
  ignored.
\item The analysis included modelling of the covariance between
  intrinsic scatter and statistical scatter of the observables, and
  the data suggested a positive correlation in the intrinsic scatter
  of $T$ and \mgas, and $T$ and $L$, but the evidence was weak. There
  was a strong and significant correlation between the scatter in
  \mgas\ and $L$, consistent with that found in hydrodynamical
  simulations. This covariance is important as it describes the
  propagation of $L$-based selection biases to biases on other
  observable quantities.
\item The PICACS framework does not provide something for nothing --
  strong degeneracies exist within PICACS which must be broken with
  informative priors on the forms of two of the three mass scaling
  relations, or with mass estimates for some of the individual clusters.
\end{itemize}

In common with the self-consistent modelling of the scaling relations
and cluster mass function of \citet{man10a,man10b}, PICACS represents a new
way of thinking about the galaxy cluster scaling relations. The PICACS
technique has many potential applications:
\begin{itemize}
\item It can be used to give robust measurement of the evolution of
  cluster scaling relations. This requires the extension of PICACS to
  incorporate selection functions, which will be the subject of a
  forthcoming paper.
\item The PICACS scaling relations for the REXCESS data provide a
  self-consistent description of that representative cluster
  population. This will allow for useful comparisons with simulated
  cluster populations -- in order to provide a good description of
  real clusters, the simulated populations should match all three of
  the PICACS scaling relations, and their covariance.
\item The PICACS framework is trivially extendable to incorporate
  additional observational data for clusters. Essentially any cluster
  observable that is expected to correlate with cluster mass
  (e.g. gravitational lensing mass estimates, Sunyaev-Zel'dovich
  effect signals, galaxy richness and dynamics) can be added to the
  framework. This will provide a natural way to test the
  self-consistency of the different cluster mass estimators, as well
  as maximise the precision of the mass constraints by combining all
  of the available information.
\end{itemize}

\section{Acknowledgements}
We thank Stefano Andreon for useful discussions of Bayesian analysis
of cluster scaling relations in general, and for providing useful
comments on a draft of this paper. We also thank Adam Mantz and
Stefano Ettori their useful comments on this work, Gabriel Pratt for
providing the REXCESS data in electronic form, and the referee for
careful reading of the manuscript and many useful comments.

\bibliographystyle{mn2e}
\bibliography{clusters}

\end{document}